\documentclass[twocolumn]{aastex63}
\usepackage{mathtools}
\usepackage[utf8]{inputenc}
\usepackage{booktabs}
\usepackage{pifont}
\newcommand{\xmark}{\ding{53}}%

\graphicspath{{./}{figures/}}

\submitjournal{AJ}

\shorttitle{HST transmission spectroscopy of WASP-79b}
\shortauthors{Rathcke et al.}

\begin{document}

\title{HST PanCET Program: A Complete Near-UV to Infrared \\ Transmission Spectrum for the Hot Jupiter WASP-79b}

\correspondingauthor{Alexander D. Rathcke}
\email{rathcke@space.dtu.dk}

\author[0000-0002-4227-4953]{Alexander D. Rathcke}
\affil{DTU Space, National Space Institute, Technical University of Denmark, Elektrovej 328, DK-2800 Kgs. Lyngby, Denmark}

\author[0000-0003-4816-3469]{Ryan J. MacDonald}
\affiliation{Department of Astronomy and Carl Sagan Institute, 
Cornell University, 122 Sciences Drive, Ithaca, NY 14853, USA}

\author[0000-0003-3726-5419]{Joanna K. Barstow}
\affil{School of Physical Sciences, The Open University, Walton Hall, Milton Keynes MK7 6AA, UK}
\affil{Department of Physics and Astronomy, University College London, Gower Street, London WC1E 6BT, UK}

\author[0000-0002-8515-7204]{Jayesh M. Goyal}
\affiliation{Department of Astronomy and Carl Sagan Institute, 
Cornell University, 122 Sciences Drive, Ithaca, NY 14853, USA}

\author[0000-0003-3204-8183]{Mercedes Lopez-Morales}
\affiliation{Center for Astrophysics $|$ Harvard {\rm \&} Smithsonian, 60 Garden Street, Cambridge, MA 02138, USA}

\author[0000-0002-6907-4476]{João M. Mendonça}
\affil{DTU Space, National Space Institute, Technical University of Denmark, Elektrovej 328, DK-2800 Kgs. Lyngby, Denmark}

\author[0000-0002-1600-7835]{Jorge Sanz-Forcada}
\affil{Centro de Astrobiolog\'{i}a (CSIC-INTA), ESAC Campus, Villanueva de la Cañada, Madrid, Spain}

\author[0000-0003-4155-8513]{Gregory W. Henry}
\affil{Center of Excellence in Information Systems, Tennessee State University, Nashville, TN  30209  USA}

\author[0000-0001-6050-7645]{David K. Sing}
\affil{Department of Earth and Planetary Sciences, Johns Hopkins University, Baltimore, MD 21218, USA}

\author[0000-0003-4157-832X]{Munazza K. Alam}
\affiliation{Center for Astrophysics $|$ Harvard {\rm \&} Smithsonian, 60 Garden Street, Cambridge, MA 02138, USA}

\author[0000-0002-8507-1304]{Nikole~K. Lewis}
\affiliation{Department of Astronomy and Carl Sagan Institute, 
Cornell University, 122 Sciences Drive, Ithaca, NY 14853, USA}

\author[0000-0002-4552-4559]{Katy L. Chubb}
\affil{SRON Netherlands Institute for Space Research, Sorbonnelaan 2, 3584 CA, Utrecht, Netherlands}

\author[0000-0003-4844-9838]{Jake Taylor}
\affil{Department of Physics (Atmospheric, Oceanic and Planetary Physics), University of Oxford, Parks Rd, Oxford, OX1 3PU, UK}

\author[0000-0002-6500-3574]{Nikolay Nikolov}
\affil{Space Telescope Science Institute, 3700 San Martin Dr, Baltimore, MD 21218, USA}

\author[0000-0003-1605-5666]{Lars A. Buchhave}
\affil{DTU Space, National Space Institute, Technical University of Denmark, Elektrovej 328, DK-2800 Kgs. Lyngby, Denmark}

\begin{abstract}
We present a new optical transmission spectrum of the hot Jupiter WASP-79b. We observed three transits with the STIS instrument mounted on HST, spanning $0.3 - 1.0~\micron$. Combining these transits with previous observations, we construct a complete $0.3 - 5.0~\micron$ transmission spectrum of WASP-79b. Both HST and ground-based observations show decreasing transit depths towards blue wavelengths, contrary to expectations from Rayleigh scattering or hazes. We infer atmospheric and stellar properties from the full near-UV to infrared transmission spectrum of WASP-79b using three independent retrieval codes, all of which yield consistent results. Our retrievals confirm previous detections of $\text{H}_2\text{O}$ (at $4.0\sigma$ confidence), while providing moderate evidence of H$^{-}$ bound-free opacity ($3.3\sigma$) and strong evidence of stellar contamination from unocculted faculae ($4.7\sigma$). The retrieved H$_2$O abundance ($\sim 1\%$) suggests a super-stellar atmospheric metallicity, though stellar or sub-stellar abundances remain consistent with present observations (O/H = $0.3 - 34 \times$ stellar). All three retrieval codes obtain a precise H$^{-}$ abundance constraint: log($X_{\rm{H^{-}}}$) $\approx -8.0 \pm 0.7$. The potential presence of H$^{-}$ suggests that JWST observations may be sensitive to ionic chemistry in the atmosphere of WASP-79b. The inferred faculae are $\sim 500~\text{K}$ hotter than the stellar photosphere, covering $\sim 15\%$ of the stellar surface. Our analysis underscores the importance of observing UV – optical transmission spectra in order to disentangle the influence of unocculted stellar heterogeneities from planetary transmission spectra.

\end{abstract}

\keywords{methods: observational – planets and satellites: atmospheres – planets and satellites: gaseous planets – methods: data analysis}

\section{Introduction} \label{sec:intro}

Transmission spectroscopy has proven a powerful method to study the atmospheres of transiting exoplanets. This technique takes advantage of the differing wavelength-dependence of absorption and scattering processes in planetary atmospheres, resulting in a wavelength-dependent planetary radius during transit \citep{Seager2000,Brown2001a}. Transmission spectra are sensitive to molecular, atomic, and ionic species, temperature structures, clouds, and hazes at the day-night terminator region \citep[see][for a recent review]{Madhusudhan2019}. If the transit chord exhibits different stellar properties from the average stellar disk, transmission spectra are also sensitive to unocculted spots or faculae \citep[e.g.][]{Rackham2018,Pinhas2018}.

The last two decades have shown \emph{Hubble Space Telescope} (HST) transmission spectroscopy observations to be very successful in probing the atmospheres of giant planets, yielding detection of several species. A non-exhaustive list of HST highlights include: detections of the alkali metals Na and K (e.g., \citealt{Charbonneau2002,Nikolov2014,Alam2018}), escaping atomic species from large exospheres (e.g., \citealt{Vidal-Madjar2003,Ehrenreich2015}), H$_2$O detections and abundance measurements (e.g., \citealt{Deming2013,Pinhas2019}), thermal inversions \citep[e.g.,][]{Evans2017,Baxter2020}, and a diverse range of cloud and haze properties \citep[e.g.,][]{Sing2016,Gao2020}. Transmission spectra of Neptune-sized and sub-Neptune-sized planets (e.g., \citealt{Crossfield2017,Benneke2019,Libby-Roberts2020}) have also been reported. Ground-based observations have also reported several detections, including Na (e.g., \citealt{Sing2012,Nikolov2018}), K (e.g., \citealt{Nikolov2016,Sedaghati2016}), Li (e.g., \citealt{Tabernero2020}), He (e.g., \citealt{Nortmann2018,Allart2018}, and clouds/hazes (e.g., \citealt{Huitson2017}). These results illustrate a dynamic movement from characterization of individual exoplanet atmospheres to a statistically significant sample. High-quality transmission spectra spanning a wide wavelength range enable precision retrievals of atmospheric properties, allowing comparative studies across the exoplanet population \citep[e.g.][]{Barstow2017,Welbanks2019b}.

Here we present a new optical transmission spectrum of the hot Jupiter WASP-79b, part of the HST Panchromatic Comparative Exoplanetary Treasury Program (PanCET)(PIs: Sing \& López-Morales, Cycle 24, GO 14767). PanCET targeted 20 planets, allowing a simultaneous ultra-violet, optical, and infrared comparative study of exoplanetary atmospheres. This program also offers valuable observations in the UV and blue-optical ($\lambda < 0.6\,\micron$) that will be inaccessible to the \emph{James Webb Space Telescope} (JWST).

WASP-79b was discovered in 2012 by the ground-based, wide-angle transit search WASP-South \citep{Smalley2012}. WASP-79b is an inflated hot Jupiter with $R_{\text{p}}$ = 1.7 $\text{R}_\text{J}$, $M_{\text{p}}$ = 0.9 $\text{M}_\text{J}$, and a mean density of $\rho$ $\sim$ 0.23 $\text{g}\text{ cm}^{-3}$. It orbits its  host star WASP-79 (also known as CD-30 1812) with a period of $P$ = 3.662 days. WASP-79 is of spectral type F5 \citep{Smalley2012} and is located in the constellation Eridanus 248 pc from Earth \citep{GaiaCollaboration2018}, making it relatively bright with $V$ = 10.1 mag. WASP-79b exhibits spin-orbit misalignment between the spin axis of the host star and the planetary orbital plane, revealing that this planet follows a nearly polar orbit \citep{Addison2013}. Recently, \citet{Sotzen2020} reported evidence for H$_2$O and FeH absorption in WASP-79b's atmosphere. They used near-infrared HST Wide Field Camera 3 (WFC3) transmission spectra observations, combined with ground-based Magellan/Low Dispersion Survey Spectrograph 3 (LDSS3) optical transmission spectra. Similar findings were reported by \citet{Skaf2020} for a different WFC3 data reduction. 

Here, we expand upon previous studies of WASP-79b's transmission spectrum by presenting new HST/STIS observations. Our paper is structured as follows. In Section \ref{sec:obs_and_reduc} we present our observations and reduction procedure. We present the analysis of the light curves in Section \ref{sec:analysis}, and assess the likelihood of stellar activity contaminating our transmission spectrum in Section \ref{sec:stellar_activity}. We then go on to describe our retrieval procedures, and present the results from these in Section \ref{retrieval0}, discuss the results in Section \ref{discussion}, and summarize in Section \ref{summary}.

\section{Observations and Data Reduction} \label{sec:obs_and_reduc}
\subsection{Observations} \label{subsec:obs}

WASP-79b was observed during three primary transit events with HST STIS, two with the G430L grating and one with the G750L grating. The specific observing dates and instrument settings are summarized in Table~\ref{tab:obsinfo}. Combined, the two gratings cover the wavelength regime from 2900 \text{\AA} to 10270 \text{\AA}, with an overlapping region from $\sim$5260-5700 \text{\AA}. The two gratings have a resolving power of $\sim$2.7 and $\sim$4.9 \text{\AA} per pixel for the G430L and G750L gratings, respectively. Thus, they offer a resolution of R $= \lambda/\Delta\lambda$ $= 500 - 1000$.

\begin{figure*}[htbp]
	\centering
	\includegraphics[width=\textwidth, trim={2.8cm 0.0cm 4.0cm 0.0cm},clip]{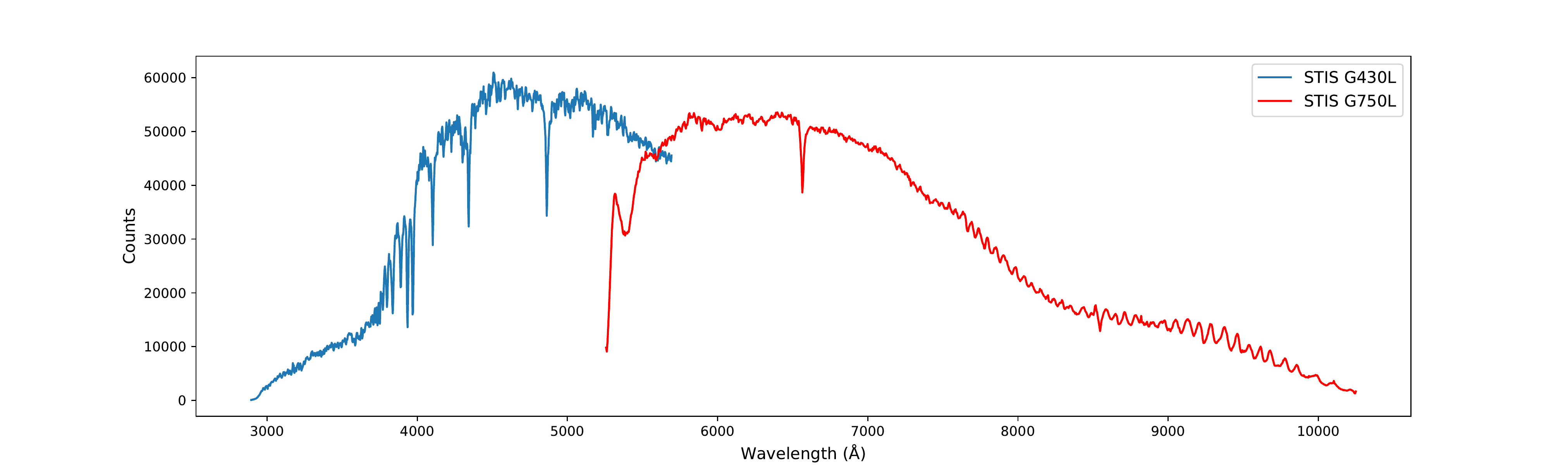}
	\caption{Sample stellar spectra of WASP-79, obtained from the STIS G430L grating (blue) and the G750L grating (red).}
	\label{fig:sample_spec}
\end{figure*}

Each transit event consists of 57-77 spectra, spanning five HST orbits, where each HST orbit takes about $\sim$95 minutes. Because HST is in a low-Earth orbit, the data collection is truncated for $\sim$45 minutes in each orbit when HST is occulted by the Earth. The observations were scheduled such that the transit event occurs in the third and fourth HST orbit, while the remaining orbits provide an out-of-transit baseline before and after each transit. All observations were made with the 52x2 $\text{arcsec}^2$ slit to minimize slit losses. Readout times were reduced by only reading out a 1024x128 pixel subarray of the CCD. This strategy has previously been found to deliver high signal-to-noise ratios (SNR) near the Poisson limit during the transit events (e.g., \citealt{Huitson2012,Sing2013}). We show an example G430L and G750L spectra of WASP-79 in Figure~\ref{fig:sample_spec}.

\begin{deluxetable*}{ccccc}
\tablecaption{HST/STIS observing information \label{tab:obsinfo}}
\tablehead{
\colhead{UT date} & \colhead{Visit number} & \colhead{Optical element}  & \colhead{\# of spectra} & \colhead{Integration time (s)} }
\startdata
2017-10-08 & 67 & G430L & 57 & 207\\
2017-10-23 & 68 & G430L & 57 & 207\\
2017-11-03 & 69 & G750L & 77 & 150$^{*}$\\
\enddata
\tablenotetext{*}{The integration times for the last orbit were only 149 s.}
\end{deluxetable*}

\subsection{Cosmic Ray Correction}
The relatively long exposure times (149-207 s) meant that our images was contaminated by multiple cosmic rays. Similar to previous studies, we found that correcting for cosmic rays using the CALSTIS\footnote{CALSTIS comprises software tools developed for the calibration of STIS data \citep{Katsanis1998} inside the IRAF environment.} pipeline did not yield satisfactory results. We, therefore, performed a custom cosmic ray correction procedure based largely on the method described by \citet{Nikolov2014}, which we explain here. For all the .flt images to be corrected, we created four difference images between the image itself and its two neighboring images on both sides in time. This effectively canceled out the stellar flux and left only the cosmic rays, which was seen as positive values for the image we were correcting and negative values for the neighboring image. Next, we created a median combined image from the four difference images, leaving only the cosmic ray events that we sought to identify and replace. For all pixels in the median image, we then computed the median of the 20 closest pixels in that row and flagged the pixel in question if it exceeded a 4$\sigma$ threshold in that window. When all pixels in an image were analyzed, we replaced all the flagged pixels by a corresponding value obtained from the four nearest images in time. All pixels flagged as `bad' by CALSTIS in the corresponding data quality frames was replaced in the same manner. Additionally, we inspected the extracted (see Section~\ref{subsec:reduc}) 1D spectra for any potential cosmic ray hits missed by the procedure applied on the .flt images. This was done by comparing every pixel with the corresponding value in all of the other spectra and flagging values more than 5$\sigma$ above the median of that pixel. We found that a few cosmic ray hits still persisted, which further investigations revealed to be located primarily close to the peak of the stellar point-spread function in the .flt images. These was corrected by replacing them in the same manner as before, by using the four nearest images in time.

\subsection{Data Reduction and Spectral Extraction} \label{subsec:reduc}

We performed a uniform data reduction for all the STIS data. The data was bias-, flat-, and dark-corrected using the latest version of CALSTIS v3.4 and the associated relevant calibration frames.

1D spectra were extracted from the calibrated and corrected .flt science frames using the APALL procedure in IRAF. To determine what aperture size to use when running the APALL procedure, a number of different widths were tested, ranging from 9 to 17 pixels with a step size of 2. The aperture
which provided the smallest out-of-transit baseline flux photometric scatter were then chosen. For all datasets, we found that this were achieved with an aperture of width 13. Like previous studies (e.g., \citealt{Sing2013}), no background subtraction was used as the
background contribution is known to have a negligible effect. Ignoring the background can even help minimize the out-of-transit residual scatter \citep{Sing2011,Nikolov2015}. The extracted spectra were then mapped to a wavelength solution obtained from the .x1d files. The discrepancy between exposure times for the G750L visit during the last HST orbit (exposure times of 149 s) compared to the preceding orbits (exposure times of 150 s) was corrected by extrapolating for the missing second under the assumption that the detector was still operating within its linearity regime.

\section{Analysis} \label{sec:analysis}

\subsection{Analysis procedure} \label{sec:ana_prodecure} 

To allow for analyses to be done in a fully Bayesian framework, all fits were carried out by: (1) treating each light curve as a Gaussian process (GP), which we implemented through use of the GP Python package \texttt{george} \citep{Ambikasaran2015}; and (2) using the Nested Sampling (NS) \citep{Skilling2004} algorithm \texttt{MultiNest} \citep{Feroz2008,Feroz2009,Feroz2019}, implemented by the Python package \texttt{PyMultiNest} \citep{Buchner2014}, which we combined with the GP likelihood function to conduct parameter inference.

GPs have been widely applied by the exoplanet community in recent years. Common applications include modeling stellar activity signals in radial velocity data (e.g., \citealt{Rajpaul2015,Jones2017}) and the correction of instrumentally induced systematics in transit data (e.g., \citealt{Gibson2012,Sedaghati2017,Evans2018}). GPs offer a non-parametric approach that finds a distribution over all possible functions that are consistent with the observed data. The main assumption behind GPs is that the input comes from infinite-dimensional data where we have observed some finite-dimensional subset of that data and this subset then follows a multivariate normal distribution. This yields the key result that input data which lie close together in input space will also produce outputs that are close together. Formally, a GP is fully defined by a mean function and a kernel (covariance) function, and it is the kernel that determines the similarity of the inputs and how correlated the corresponding outputs are. Given the rapidly growing and already extensive use of GPs applied in the literature, we refer readers unfamiliar with GPs and their applications to this type of analysis
to \citet{Gibson2012}, which gives a good introduction to their uses on transmission spectroscopy data. 

We adopted the analytic transit model of \citet{Mandel2002} for our GP mean function. This model is a function of mid-transit times ($t_0$), the orbital period ($P$), the planet-star radius ratio ($R_p/R_\star$), the semi-major axis in units of stellar radii ($a/R_\star$), the orbital inclination ($i$), and limb darkening coefficients. We used the \texttt{BATMAN} package to implement the transit model \citep{Kreidberg2015}. Uncertainties for each data point were initially derived based solely on Poisson statistics.

Nested sampling is a numerical method for Bayesian computation targeted at efficient calculation of the Bayesian evidence, with posterior samples produced as a by-product. Compared to traditional MCMC techniques, NS is able to sample from multi-modal and degenerate posteriors efficiently. It is a Monte Carlo algorithm that explores the posterior distribution by initially selecting a set of samples from the prior, called live points. The live points are then iteratively updated by calculating their individual likelihoods and replacing the live point with the lowest likelihood. This procedure ensures an increasing likelihood as the prior volume shrinks through each iteration and runs until a specified tolerance level is achieved. 

\subsubsection{Model Comparison} \label{sec:model_comparison}

Our overall approach offers several advantages, some of which we briefly highlight here. Rather than enforcing a parametrized function to model the systematic effects, the GP allows for a non-parametric approach that simultaneously fits for both the transit and systematics. 
Picking an optimal systematics model or, in our case, an optimal kernel function requires the conduction of model comparison.
It is a general problem that such optimization routines can lead to overfitting. This problem has well-established solutions, such as the Bayesian information criterion (BIC) \citep{Schwarz1978} and the Akaike information criterion (AIC) \citep{Akaike1974}. These two corrective terms to maximum likelihoods both enforce a penalty, based on the number of model parameters, but are slightly different in the way they introduce the penalty term. One potential critical flaw of the BIC and AIC approaches is that the model selection is based on a single maximum likelihood estimate, which does not consider the uncertainties of the model parameters, $\pmb{\theta}$. 
Rather than relying on these methods, our application of NS allowed us to not only carry out posterior inference but also model comparison in a Bayesian framework. The model comparison was done as follows: With $\pmb{\theta}$ being the parameter vector, $\pmb{D}$ the data, and $\mathcal{M}$ the model, then Bayes' theorem is given by

\begin{equation}\label{eq:bayes_1}
    p(\pmb{\theta}|\pmb{D},\mathcal{M}) = \frac{p(\pmb{D}|\pmb{\theta},\mathcal{M})p(\pmb{\theta}|\mathcal{M})}{p(\pmb{D}|\mathcal{M})},
\end{equation}
\noindent
where $ p(\pmb{\theta}|\pmb{D},M)$ is the posterior probability distribution for $\pmb{\theta}$, $p(\pmb{D}|\pmb{\theta},M)$ (hereafter $\mathcal{L}(\pmb{\theta})$) the likelihood, $p(\pmb{\theta}|M)$ (hereafter $\pi(\pmb{\theta})$) the prior probability, and $p(\pmb{D}|M)$ (hereafter $\mathcal{Z}$) is called the evidence or marginal likelihood.

$\mathcal{Z}$ is a normalization constant for the posterior and is computed from samples produced from the posterior probability distribution of $\pmb{\theta}$ as

\begin{equation}\label{eq:bayes_2}
    \mathcal{Z} = \int \mathcal{L}(\pmb{\theta})\pi(\pmb{\theta})d\pmb{\theta}.
\end{equation}
\noindent
It follows that the posterior probability of model $\mathcal{M}$ is

\begin{equation}\label{eq:bayes_3}
    p(\mathcal{M}|\pmb{D}) = \frac{p(\pmb{D}|\mathcal{M})p(\mathcal{M})}{p(\pmb{D})}.
\end{equation}
\noindent
To perform a relative comparison between two models we then took the ratio of the model posterior probabilities and cancelling the term $p(\pmb{D})$, yielding

\begin{equation}\label{eq:bayes_4}
    \frac{p(\mathcal{M}_{i}|\pmb{D})}{p(\mathcal{M}_{j}|\pmb{D})} = \frac{p(\pmb{D}|\mathcal{M}_{i}) \, p(\mathcal{M}_{i})}{p(\pmb{D}|\mathcal{M}_{j}) \, p(\mathcal{M}_{j})} = \frac{\mathcal{Z}_{i} \, \pi(\mathcal{M}_{i})}{\mathcal{Z}_{j} \, \pi(\mathcal{M}_{j})}.
\end{equation}
\noindent
With no \textit{a priori} model preferences the $\pi(\mathcal{M}_{i}) / \pi(\mathcal{M}_{j})$ term cancels out, leaving us with only the evidence ratio $\mathcal{Z}_{i} / \mathcal{Z}_{j}$. This ratio is commonly referred to as the Bayes factor \citep{Kass1995} and is what we used to directly compare two models. This model comparison comes with the benefit of incorporating Occam's razor, automatically penalizing unreasonable model complexity that in turn would lead to overfitting (and worse predictive power). Hence, our approach eliminates the need for methods such as BIC or AIC to help perform kernel choices. However, we note that the evidence calculation is based not only on the choice of kernel, but also on the optimization of the hyperparameters, which can potentially get caught in bad local optima. This is usually accounted for by running the optimization routine multiple times with different starting conditions for the hyperparameters (see e.g., the Mauna Loa atmospheric $\text{CO}_2$ example in Chapter 5 of \citealt{Rasmussen2006}), but the additional benefit of utilizing NS is its ability to handle irregular likelihood surfaces while still being efficient compared to MCMC methods. While this does not guarantee finding the global optima, we chose to rely on its ability to handle such a likelihood surface, as this provided us with a significant computational speed-up.

\subsubsection{Kernel Selection} \label{sec:kernel_selection}

With a way of evaluating the comparative performance between kernels, we set out to determine what kernel to use in the light curve fits. We used time as input variable and included the following five different `standard' kernels in this investigation:

\begin{enumerate}
    \item Squared Exponential: 
        \begin{equation}\label{eq:squaredexp_kernel}
            k(x_{n},x_{m}) = \sigma^{2} \exp\left(-\frac{(x_{n}-x_{m})^{2}}{2\ell^{2}}\right)
        \end{equation}
    \item Rational Quadratic:
        \begin{equation}\label{eq:RQ_kernel}
            k(x_{n},x_{m}) = \sigma^{2}\left(1+\frac{(x_{n}-x_{m})^{2}}{2\alpha\ell^{2}}\right)^{-\alpha}
        \end{equation}
    \item Mátern 3/2:
        \begin{align}\label{eq:matern32kernel}
            k(x_{n},x_{m}) = \sigma^{2} 
            & \left(1+\frac{\sqrt{3|x_{n}-x_{m}|^{2}}}{\ell}\right) \times \nonumber \\
            &\hspace{-8pt} \exp \left(-\frac{\sqrt{3|x_{n}-x_{m}|^{2}}}{\ell}\right)
        \end{align}
    \item Periodic:
        \begin{equation}\label{eq:per_kernel}
            k(x_{n},x_{m}) = \sigma^{2} \exp\left(-\frac{2 \sin^{2}(\pi|x_{n}-x_{m}|/p)}{\ell^{2}}\right)
        \end{equation}
    \item Linear:
        \begin{equation}\label{eq:lin_kernel}
            k(x_{n},x_{m}) = \sigma^{2}\left(x_{n}-c\right)\left(x_{m}-c\right)
        \end{equation}
\end{enumerate}

where $x_{n},x_{m}$ refer to the elements in the covariance matrix, $\sigma$ is the maximum variance allowed, $\ell$ is the characteristic length scales, $\alpha$ is the Gamma distribution parameter, $p$ is the period between repetitions, and $c$ is a constant term. In addition, we also incorporated a white noise kernel in all fits, which has the form:

\begin{equation}\label{eq:whitenoise_kernel}
    \text{White Noise:   } k(x_{n},x_{m}) = \sigma_{wn}^{2}\delta(x_{n},x_{m}),
\end{equation}

\noindent where $\sigma_{wn}$ is the amplitude of the white noise (i.e., photon noise) for each data point, and $\delta(x_{n},x_{m})$ is the Kronecker delta function.

To decide which kernel to use, we tried all these kernels separately in the white light curve fits (see Section~\ref{subsec:WL_fits}) and compared them by their evidence. We then expanded upon this by utilizing the fact that any additive or multiplicative combination of these five kernels are still valid kernels. This allowed us to construct more complex kernels built from these `standard' kernels, which offered the possibility to model many different properties that some of the `standard' kernels would struggle with. Rather than enforcing the structural form of our kernel of choice, we performed a comprehensive test of different kernels in the white light curve fits. This search was carried out by setting up a grid consisting of the above-mentioned five kernels and then trying all two-component additive and multiplicative combinations. Following this step, we allowed once again for an additional `standard' kernel to be added or multiplied onto the existing two-component kernels. At this stage, we kept only the best performing kernel and attempted to expand the remaining kernel even further. We found the evidence did not improve (slightly worsened, in fact), indicating that no (or very little) structure was left. This was further reinforced by investigating the residuals after adding the 3rd `standard' kernel, which was well-described by a normal distribution with a standard deviation similar to the photon noise.

\begin{figure*}[htbp]
	\centering
	\includegraphics[width=\textwidth, trim={2.0cm 0.2cm 4.0cm 2.0cm},clip]{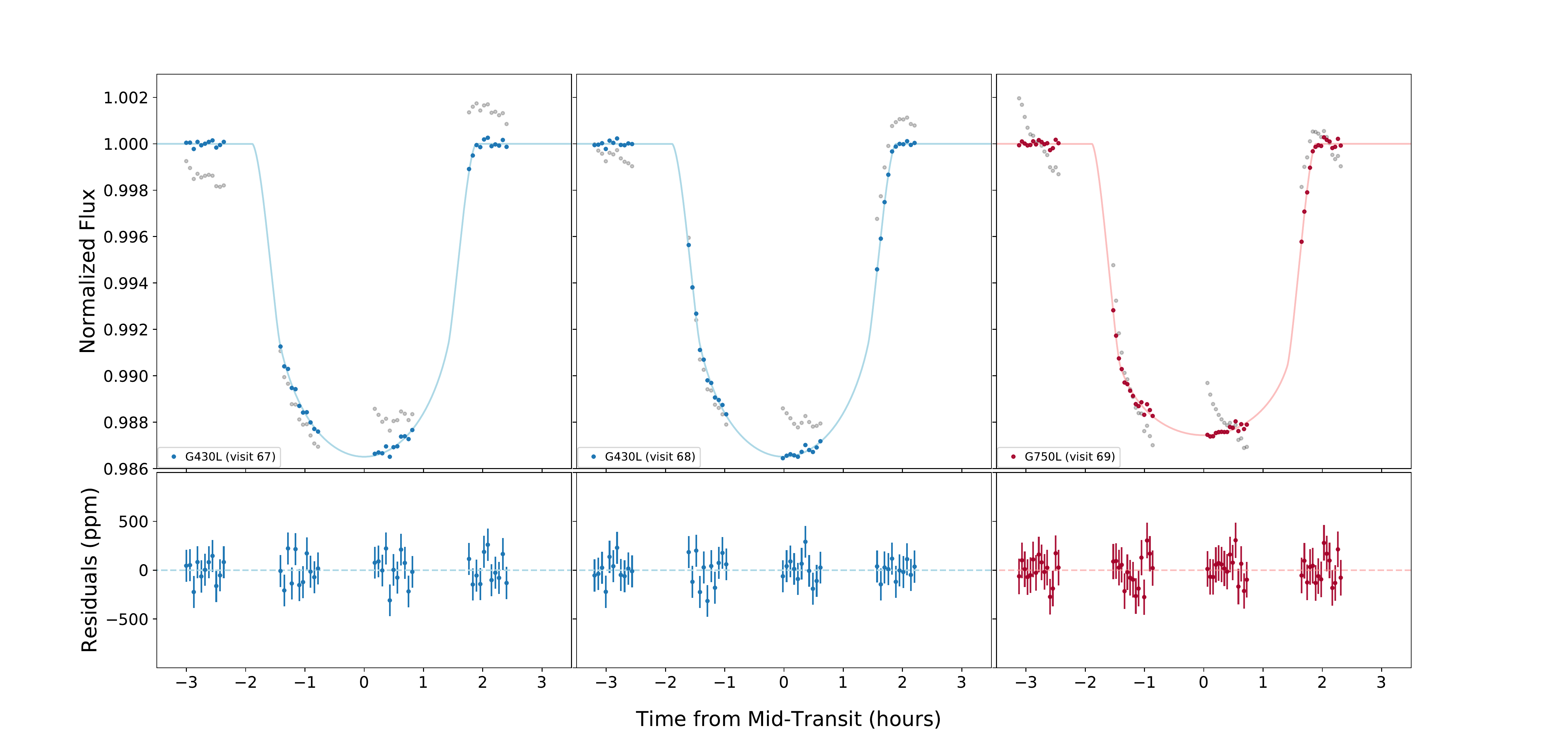}
	\caption{WASP-79b HST/STIS normalized white light curves from the data obtained during the 3 visits (left to right): visit 67 (G430L), visit 68 (G430L), and visit 69 (G750L). Top row: Data points after removing the systematic effects inferred from the GP analyses (blue and red points for the G430L and G750L gratings, respectively), with the best-fit model (solid lines) and the raw light curve data points prior to the GP analyses (transparent grey points). Bottom row: Corresponding O-C residuals, with photon noise error bars.}
	\label{fig:w79_wl}
\end{figure*}

\subsection{Limb Darkening Treatment}\label{sec:limb_darkening}
The treatment of stellar limb darkening effects can have a significant impact on derived transmission spectra. Optimally, these effects could be accounted for by fitting for the coefficients defining a given limb darkening model. However, the incomplete phase coverage and the relatively low temporal sampling rate of the observations make it difficult to derive the coefficients directly from the data. Instead, we used the \texttt{Limb Darkening Toolkit (LDTk)} \citep{Parviainen2015} Python package, which utilizes the PHOENIX stellar models of \citet{Husser2013} to calculate limb darkening coefficients. This procedure allowed us to fix (rather than fit) the limb darkening coefficients in our light curve models to theoretical values, which gave us the advantage of freely picking a limb darkening law parametrized by a higher number of coefficients. Therefore, we made use of the non-linear limb darkening law described by four parameters \citep{Claret2003} and estimated the parameters based on the PHOENIX stellar model grid point closest to that of WASP-79. The resulting coefficients used in the light curve fits are shown in Table~\ref{tab:tsres79}.

\subsection{White Light Curve Fits} \label{subsec:WL_fits}

To refine system parameters for the planet, we initially performed fits for the light curves produced by a summation of the entire dispersion axis, commonly referred to as a white light curve. As several of the physical parameters of the system are wavelength-independent, we unsurprisingly obtain the most precise system parameters when including the entire wavelength range as this ensured the highest possible SNR.

In accordance with common practice, we discarded all exposures from the first HST orbit and the first exposure of each subsequent orbit as they are known to suffer from unique and complex systematics arising from the telescope thermally relaxing into its new pointing position (\citealt{Brown2001b}). We conducted the white light curve fit jointly for the two G430L grating visits, but separately from the G750L grating visit. Furthermore, we assumed a circular orbit (zero eccentricity), in correspondence with the results of \citealt{Smalley2012}, for all light curve fits. This enabled us to perform our fits by allowing $t_0$, $R_p/R_\star$, $a/R_\star$, $i$, and the hyperparameters related to the kernel function of the GP to vary as free parameters.

As described above (see Section~\ref{sec:ana_prodecure}) we set out to find a suitable kernel for the GP, and this search resulted in a composite kernel consisting of the white noise kernel, a periodic kernel times a squared exponential kernel (hereafter referred to as a locally periodic kernel) and a Matérn-3/2 kernel. The resulting multi-component kernel takes the form:

\begin{align}\label{eq:composite_kernel}
    k(x_{n},x_{m}) & = \sigma_{wn}^{2}\delta(x_{n},x_{m}) \nonumber \\ 
    &\hspace{-45pt} + \sigma_{a}^{2} \exp\left(-\frac{2 \sin^{2}(\pi|x_{n}-x_{m}|/p)}{\ell_{a}^{2}}\right) \exp\left(-\frac{|x_{n}-x_{m}|^{2}}{2\ell_{b}^{2}}\right) \nonumber \\
    &\hspace{-28pt} + \sigma_{b}^{2}\left(1+\frac{\sqrt{3|x_{n}-x_{m}|^{2}}}{\ell_{c}}\right) \exp\left(-\frac{\sqrt{3|x_{n}-x_{m}|^{2}}}{\ell_{c}}\right)
\end{align}

\noindent
where $\sigma_{a}, \sigma_{b}$ and $\ell_{a},\ell_{b},\ell_{c}$ are the allowed variance and correlation length scales for the corresponding part of the composite kernel.

We note that while the search for the structural form of the kernel was determined by the data itself, the individual components of the kernel will reflect fits to physically introduced systematic effects (which could be of instrumental or astrophysical origin). Here, the locally periodic kernel component is physically motivated by the well-known breathing effect, which introduces substantial correlated systematics in the data \citep{Brown2001b}. This effect is the product of the spacecraft suffering from significant thermal variations in its low Earth $\sim$95 minute orbits. The Matérn-3/2 kernel component is a flexible kernel, which performs well in modeling correlations on shorter length-scales, and thus are implemented to deal with residual correlated systematic effects of unknown origin. The fits assume uniform priors on all parameters. For the free transit parameters (i.e., the parameters of the GP mean function) we applied a bound on our uniform priors at $\pm$ 20$\sigma$ from the Transiting Exoplanet Survey Satellite (TESS) \citep{Ricker2014} inferred values of \citet{Sotzen2020}. For the GP kernel parameters, we set the length scale prior lower limit at zero and the upper limit at the value corresponding to the time between the first and the last observation, and the amplitude parameters were only restricted to not be larger than the difference between the minimum and maximum flux measurements. 

\begin{deluxetable}{ccc}
    \tablecaption{System Parameter Results \label{tab:sysp_w79}}
    \tablehead{
    \colhead{ } & \colhead{$a/\text{R}_\star$} & \colhead{Inclination [$^{\circ}$]}}
    \startdata
    G430L white light fit & $7.31\pm0.06$ &  $86.012\pm0.122$\\
    G750L white light fit & $7.28\pm0.09$ & $85.900\pm0.172$\\
    TESS & $7.29\pm0.08$ & $85.929\pm0.174$\\
    Weighted average & $7.29\pm0.04$ & $85.963\pm0.086$\\
    \enddata
\end{deluxetable}

The results of these fits are summarized in Table~\ref{tab:sysp_w79} and visualized in Figure~\ref{fig:w79_wl}. We found that our inferred wavelength-independent system parameters resulted in better fits as well as lower standard deviations than those quoted in the discovery paper of \citet{Smalley2012}. We, therefore, chose to use these and the TESS photometry inferred values to calculate weighted average values (also shown in Table~\ref{tab:sysp_w79}), which we use in the wavelength-binned fits.

\subsection{Spectrophotometric Light Curve fits} \label{subsec:binned_fits}

In order to assemble the transmission spectrum, we extracted light curves from wavelength bins for both gratings. Specifically, we produced wavelength-binned light curves by dividing the spectra into bins varying in size from 85 to 1000 \text{\AA}. We chose to customize bin sizes based on the criteria that the SNR in each bandpass was sufficiently high not to be dominated by photon noise, yet still small enough to preserve valuable information from the underlying transmission spectra. This criterion was achieved in all bins with an average SNR of $\sim$1500. Additionally, we also made sure that the borders of the channels did not coincide with prominent stellar lines. Fits were then carried out in each spectrophotometric channel similar to the white light curve fits (see Section~\ref{subsec:WL_fits}), but with the exception that we froze each wavelength-independent parameter to the weighted average values quoted in Table~\ref{tab:sysp_w79}. Effectively, this meant that the fits carried out in the spectrophotometric channels only allowed for the parameter of interest, $\text{R}_\text{p}/\text{R}_\star$, and the GP kernel parameters to vary as free parameters. Identical to the white light curve fits, we jointly fit the wavelength-binned light curves from our two data sets obtained with the G430L grating.

\begin{figure*}[htbp]
\hspace{-1.6cm}
\includegraphics[trim={0 1.7cm 0 2.2cm},clip]{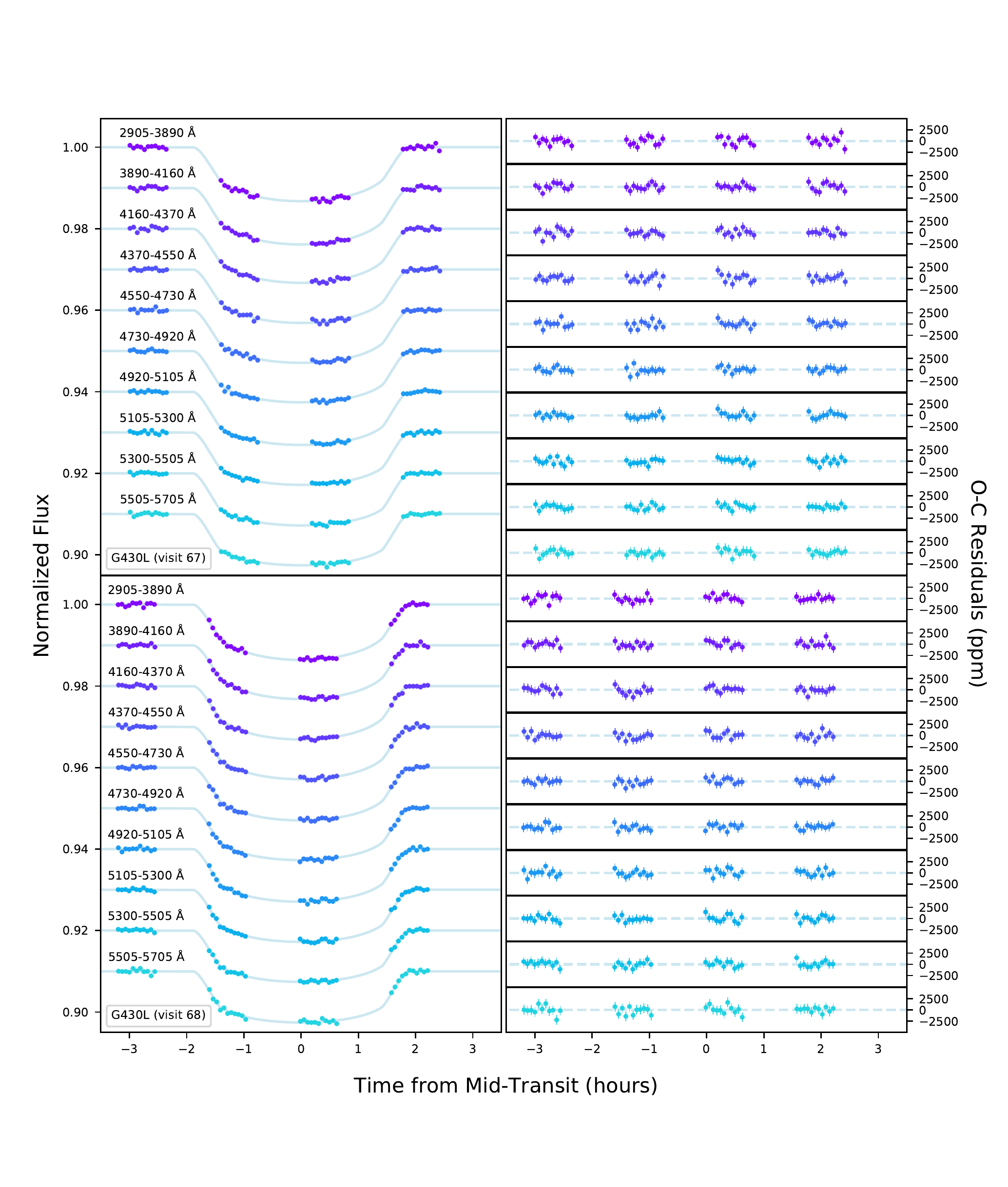}
\caption{WASP-79b HST/STIS observations from data obtained during visit 67 (top) and visit 68 (bottom) with the G430L grating. Left panel: Detrended light curves (points) and best-fit transit model (solid lines). The wavelength-binned light curves are shifted vertically by an arbitrary constant for clarity and are arranged with the bluest spectrophotometric channel on top and the reddest channel on bottom. Right panel: Corresponding O-C residuals in parts per million.}
\label{fig:w79_g430l_vstacked}
\end{figure*}

\begin{figure*}[htbp]
\hspace{-1.6cm}
\includegraphics[trim={0 1.2cm 0 2.4cm},clip]{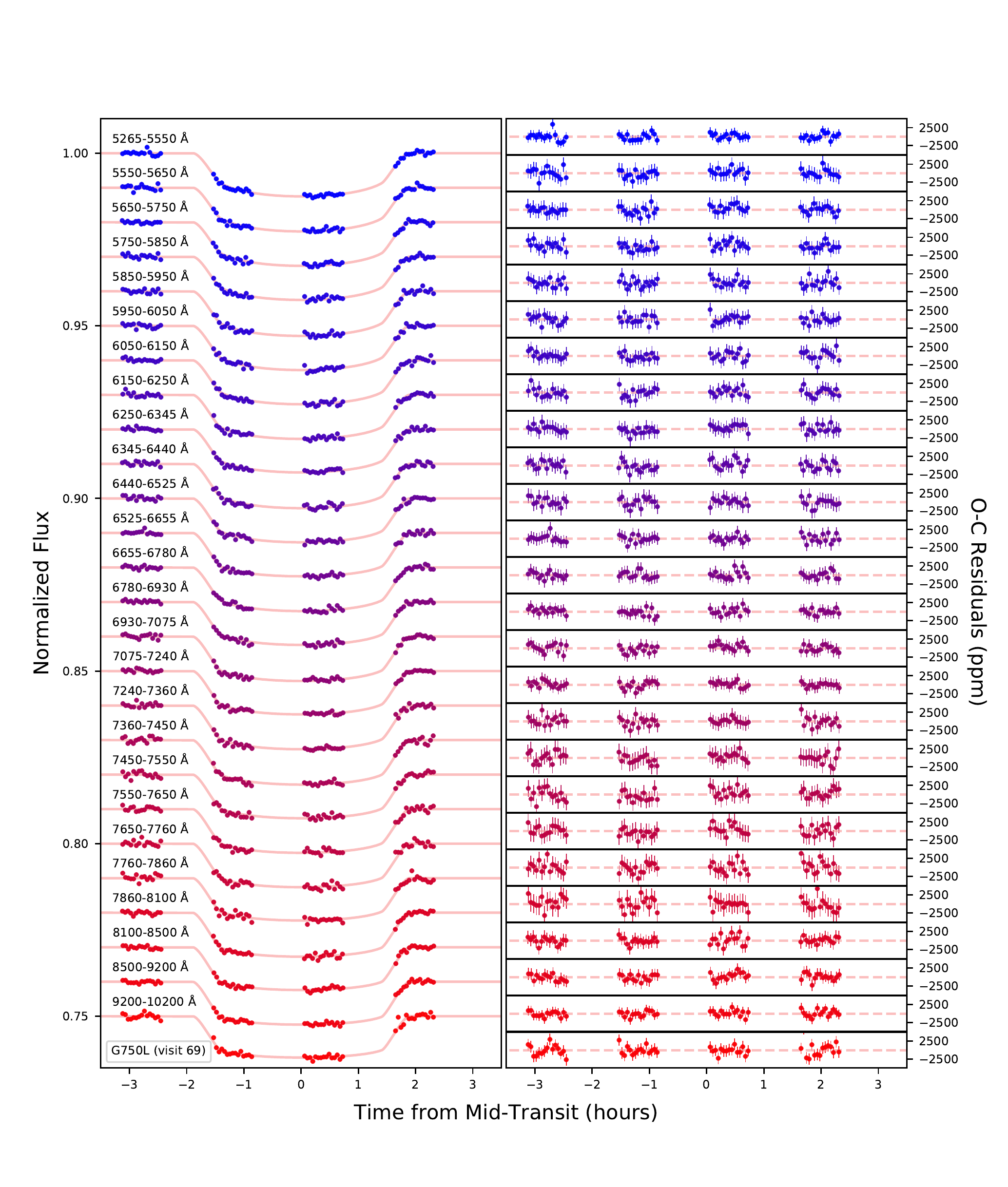}
\caption{Same as Figure~\ref{fig:w79_g430l_vstacked}, but for visit 69 with the G750L grating.}
\label{fig:w79_g750l}
\end{figure*}

\begin{figure*}[ht!]
    \centering
    \includegraphics[width=\textwidth, trim={0.0cm 0.0cm 0.0cm 0.0cm}]{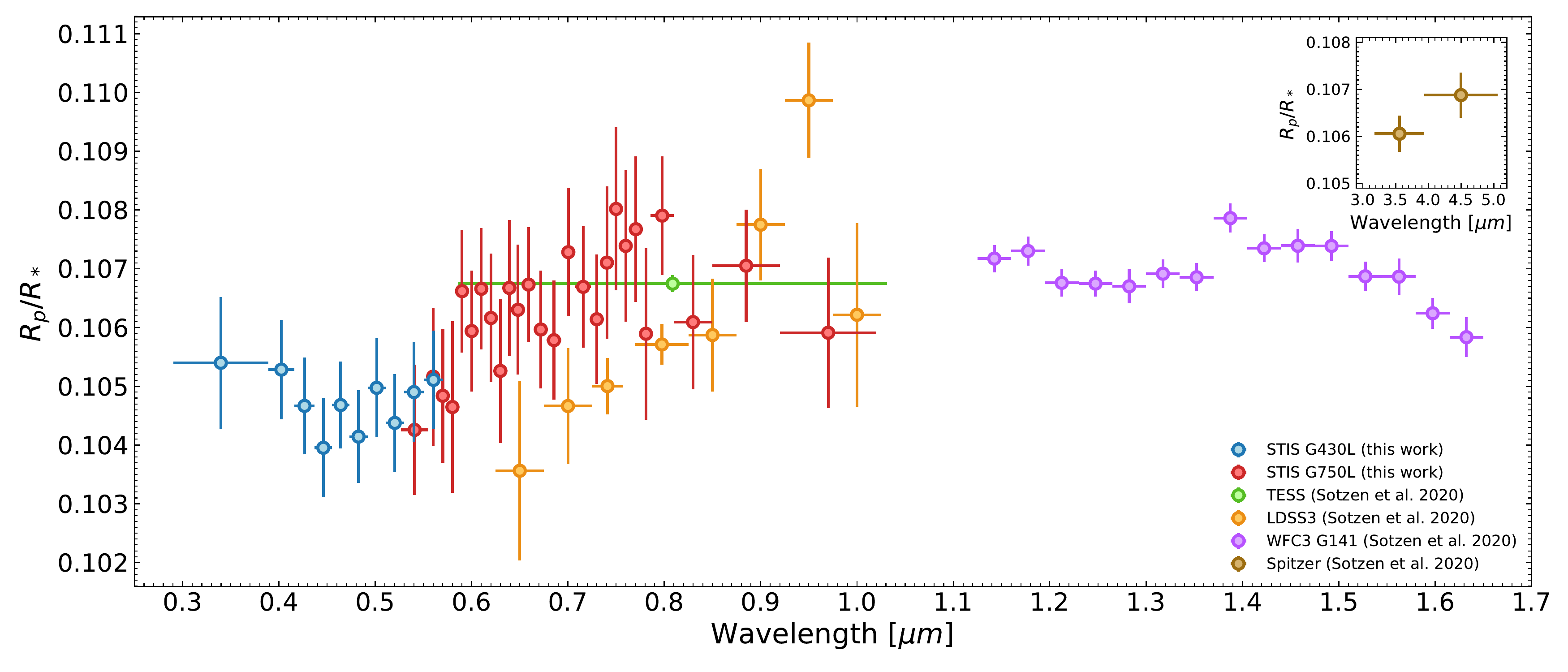}
    \caption{The HST/STIS transmission spectrum of WASP-79b. The 1$\sigma$ error bars are obtained from our posterior samples (vertical lines), with the spectrophotometric channel size (horizontal lines) (blue and red circles for G430L and G750L, respectively). Also included are observations from \citet{Sotzen2020} (green, orange, purple, and brown circles for TESS, LDSS3, WFC3, and Spitzer, respectively).} 
    \label{fig:transmission_spectrum}
\end{figure*}

We accounted for potential systematic discrepancies in the absolute transit depths between the G430L and G750L gratings and put the combined STIS transmission spectrum on an absolute scale. First, we measured the offset between the two gratings using their overlapping region between 0.53-0.57 $\mu$m. We did this by fitting the light curve jointly for the two G430L data sets and separately for the G750L data set in the overlapping region. Secondly, TESS observed 12 transits of WASP-79b in January and February of 2019, yielding a tight constraint of $\text{R}_\text{p}/\text{R}_\star = 0.10675\pm0.00014$ \citep{Sotzen2020}, that we utilized to calibrate the transmission spectrum to an absolute scale. Therefore, we performed a similar fit for the G750L grating in the $0.59-1.02~\mu\text{m}$ range corresponding to the TESS bandpass. Finally, we stitched the combined transmission spectra together by uniformly offsetting the G430L transmission spectra by the difference between the inferred values for the two gratings in the same bandpass, and then offsetting the entire transmission spectrum in the same way, anchoring it to the TESS value. The inferred values are noted in Table~\ref{tab:stitch_values}. The detrended binned light curves for all three visits are shown in Figure~\ref{fig:w79_g430l_vstacked} for visits 67 and 68, and Figure~\ref{fig:w79_g750l} for visit 69. To check for the sensitivity of our limb darkening treatment, we repeated the analysis but applied the quadratic limb darkening law instead and fit for the two coefficients in the light curve models. We found the two treatments showed excellent consistency in the relative transit depths, and measurements in all channels agreed within 1$\sigma$ (see the Appendix, Figures~\ref{fig:qd_vs_nl_g430l} and \ref{fig:qd_vs_nl_g750l}). Our final stitching-corrected transmission spectrum is presented in Figure~\ref{fig:transmission_spectrum} (alongside the observations from \citealt{Sotzen2020}) and summarized in Table~\ref{tab:tsres79}.

A visual inspection of the transmission spectrum shows no obvious signs of absorption from sodium or potassium, but it does show decreasing transit depths towards blue wavelengths over the optical spectral range. This morphology also appears in ground-based LDSS3 data from \citet{Sotzen2020}, though those data appear systematically vertically offset from the STIS data. It is not clear what is causing this vertical offset, but some explanations include: instrumental systematics, stellar variability, different orbital parameters, and/or different limb darkening coefficients. Nevertheless, we verified that excluding the LDSS3 data does not alter our atmospheric inferences in later sections.

\begin{deluxetable}{ccc}
    \tablecaption{Stitching Parameters \label{tab:stitch_values}}
    \tablehead{
    \colhead{Data} & \colhead{$\text{R}_\text{p}/\text{R}_\star$} & \colhead{Bandpass} }
    \startdata
    G430L & 0.10519 & 0.53-0.57 $\mu$m\\
    G750L & 0.10482 & 0.53-0.57 $\mu$m\\
    G750L & 0.10662 & 0.59-1.02 $\mu$m\\
    TESS  & 0.10675 & 0.59-1.02 $\mu$m\\
    \enddata
\end{deluxetable}

\section{Stellar Activity} \label{sec:stellar_activity}

Stellar activity in the form of bright and dark spots can potentially introduce spurious features in the transmission spectra of exoplanets (e.g., \citealt{Pont2013,Mccullough2014,Rackham2018}). Therefore, we performed an extensive inspection of available observations of the star to evaluate the effect that stellar activity might have on the observed transmission spectrum. In the case of WASP-79b, its host is an F5 star with a ${\rm log~g}$ = 4.20 $\pm$ 0.15 ${\rm cgs}$, suggesting that the star is either still on the main sequence or slightly evolved \citep{Smalley2012}. Photometric time-series observations with TESS suggest the star is quiet, with no obvious signs of periodic activity, as the baseline varies within 1$\sigma$ at less than 1 mmag \citep{Sotzen2020}. 

Furthermore, \citet{Sotzen2020} included photometric observations of WASP-79 obtained with the Tennessee State University C14 Automated Imaging Telescope (AIT) at Fairborn Observatory \citep{Henry1999,Oswalt2003}. These included the 2017 and 2018 observing seasons, as well as the partial 2019 season available at the time. These observations did not reveal any significant variability within each season, nor did they indicate any significant year-to-year variability. We extend these observations by including the remainder of the 2019 observing season (adding 30 new observations). The AIT observations and their reduction are described in \citet{Sing2015}, with the complete 2019 observing season shown in the supplementary material. These observations show no obvious signs of activity, in agreement with the findings of \citet{Sotzen2020}.

From the spectroscopic observations in the WASP-79b discovery paper \citep{Smalley2012}, the residuals in the radial velocity variations of WASP-79 and the lack of a correlation between radial velocity variations and line bisector spans also suggest low levels of stellar activity. However, the star has a projected rotational velocity of ${v\rm{sin}i }$ = 19.1 $\pm$ 0.7 ${\rm km~s^{-1}}$, corresponding to a maximum rotation rate of 4.0 $\pm$ 0.8 days.

\begin{figure}
    \begin{center}
   \includegraphics[width=\columnwidth, trim={2.0cm 1.5cm 3.0cm 1.0cm},clip]{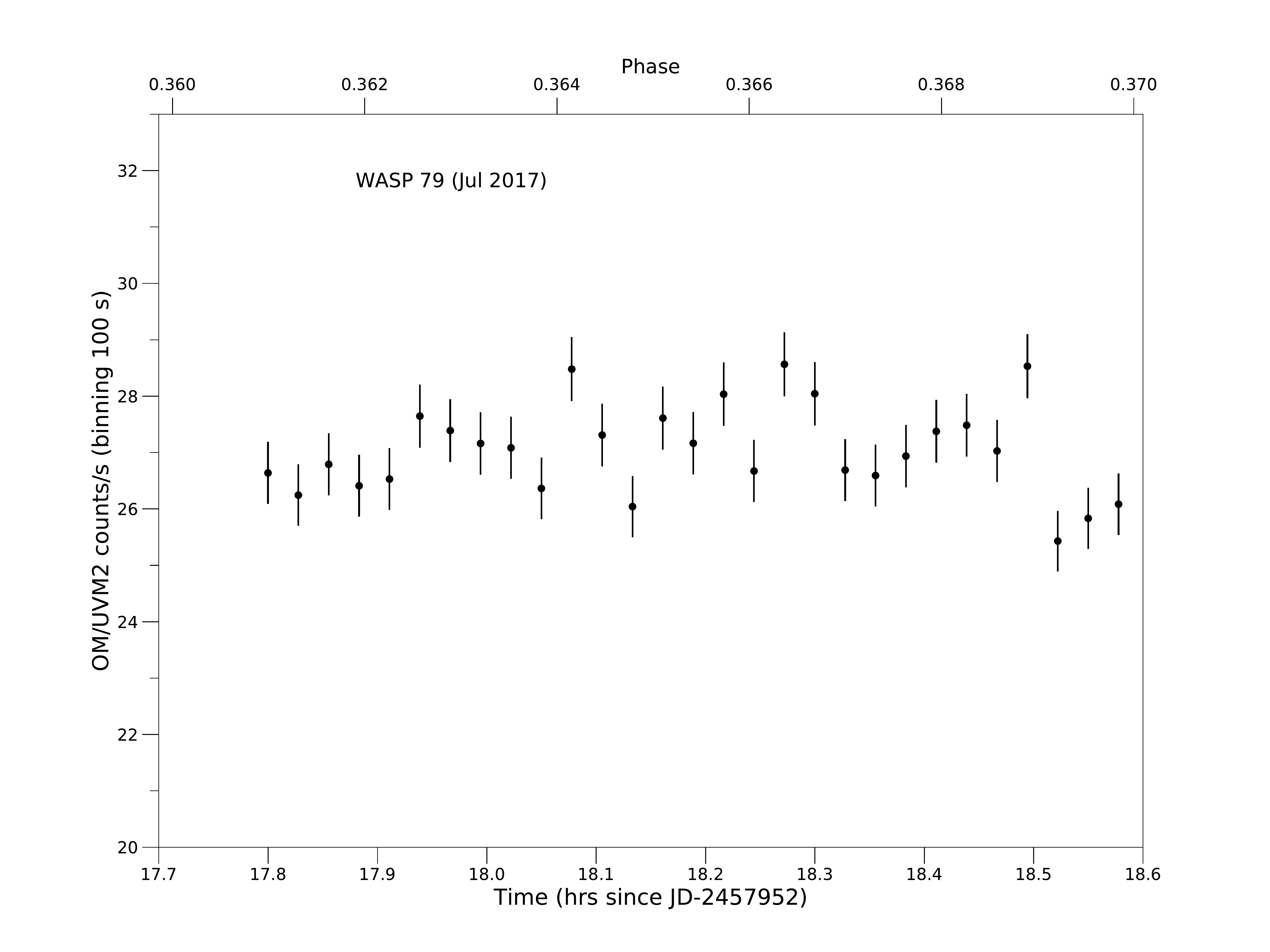}
    \caption{XMM-Newton Optical Monitor (UVM2 filter, 100 s binning) light curve. Orbital phase of WASP-79b is indicated in the upper axis, using the orbital parameters of \citet{Smalley2012}.} 
    \label{fig:xmm_lc}
     \end{center}
\end{figure}

We also considered XMM-Newton observations taken on 2017 July 18 (PI J. Sanz-Forcada) to evaluate the activity level of the star. XMM-Newton simultaneously observes with the EPIC X-ray detectors and the Optical Monitor (OM). The star was detected in X-rays (SNR=3.4) with a luminosity of $6\times 10^{28}$~erg\,s$^{-1}$ (Sanz-Forcada et al. in prep.). This implies a value of $\log L_{\rm X}/L_{\rm bol} = -5.5$, indicating a moderate level of activity \citep{2011ApJ...743...48W}. The analysis of the variability in the X-ray light curve is inconclusive, given the large error bars. However, the UV observations from XMM-Newton/OM (using the UVM2 filter, $\lambda_c=2310$~\AA) are suggestive of variability (Figure \ref{fig:xmm_lc}). Though a detailed accounting of UV variability is beyond the scope of this work, we conducted a Bayesian model comparison with the UltraNest package \citep{Buchner2021} to quantify the evidence for variability. We found a Bayes factor of 21 in preference of a sinusoidal function over a flat line (equivalent to $3\,\sigma$ evidence). The variability we infer is likely related to active regions lying in the chromosphere of the star. Although stellar activity is uncommon among early F stars, the fast rotation rate of WASP-79 and a stellar radius as high as 1.9~$R_\odot$ \citep{Smalley2012} could result in some level of activity, as has been observed in Procyon \citep[F4IV-V, $R=2.06~R_\odot$][and references therein]{Sanz-Forcada2003}. Based partly on the signs of activity in the observed UV light curve, we include the effect of starspots and faculae in the atmospheric retrieval analyses that follow.

\section{Atmospheric retrieval analysis of WASP-79b's transmission spectrum} \label{retrieval0}

We now turn to extract the planetary atmosphere and stellar properties from the transmission spectrum of WASP-79b. We employ the technique of atmospheric retrieval, which leverages a Bayesian framework to conduct parameter estimation and model comparison. This allows statistical constraints to be placed on the abundances of atomic and molecular species, the temperature structure, and the proliferation of clouds. We further include a parametrization of stellar heterogeneity to account for potential unocculted starspots or faculae.

In what follows, we first describe our modeling and retrieval approach. We then present our combined inferences concerning the atmosphere of WASP-79b and the heterogeneity of its host star. 

\newpage

\subsection{Atmospheric Retrieval Approach} \label{retrieval}

We conducted a series of atmospheric retrievals using three different codes. Each code was free to choose its own set of molecular, atomic, and ionic opacities, along with a pressure-temperature (P-T) profile and cloud / haze parametrization. Our approach, considering multiple independent retrieval codes, ensures robust atmospheric inferences. The configurations used by each retrieval code are summarized in Table~\ref{tab:retrieval_config}.

\begin{deluxetable*}{lccc} \label{tab:retrieval_config}
\tablecaption{Atmospheric Retrieval Configurations}
\tablewidth{0pt}
\tablehead{
Retrieval feature & POSEIDON & NEMESIS & ATMO 
}
\startdata \\[-8pt]
\textbf{Chemical Species} & -- & -- & -- \\ 
\hspace{0.5em} H$_2$O & \checkmark$^{1}$ & \checkmark$^{1}$ & \checkmark$^{2}$ \\
\hspace{0.5em} CO & \checkmark$^{3}$ & \checkmark$^{3}$ & \xmark \\
\hspace{0.5em} CO$_2$ & \checkmark$^{4}$ & \checkmark$^{5}$ & \xmark \\
\hspace{0.5em} CH$_4$ & \checkmark$^{6}$ & \xmark & \xmark \\
\hspace{0.5em} HCN & \checkmark$^{7}$ & \xmark & \xmark \\
\hspace{0.5em} NH$_3$ & \checkmark$^{8}$ & \xmark & \xmark \\
\hspace{0.5em} H$^{-}$ & \checkmark$^{9}$ & \checkmark$^{9}$ & \checkmark$^{9}$ \\
\hspace{0.5em} H & \checkmark$^{9}$ & \checkmark$^{9}$ & \xmark \\
\hspace{0.5em} e$^{-}$ & \checkmark$^{9}$ & \checkmark$^{9}$ & \xmark \\
\hspace{0.5em} Na & \checkmark$^{10}$ & \xmark & \xmark \\
\hspace{0.5em} K & \checkmark$^{10}$ & \xmark & \xmark \\
\hspace{0.5em} Fe & \checkmark$^{10}$ & \xmark & \xmark \\
\hspace{0.5em} TiO & \checkmark$^{11}$ & \checkmark$^{11}$ & \xmark \\
\hspace{0.5em} VO & \checkmark$^{12}$ & \checkmark$^{12}$ & \xmark \\
\hspace{0.5em} FeH & \checkmark$^{13}$ & \xmark & \xmark \\
\midrule
\textbf{P-T Profile} & \citet{Madhusudhan2009} & \citet{Guillot2010} & Isotherm \\ 
\midrule
\textbf{Clouds \& Hazes} & Patchy cloud + haze & Cloud or haze slab & Cloud + haze \\ 
\midrule
\textbf{Radius Ref. Pressure} & 10\,bar & 10\,bar & 10$^{-3}$\,bar \\ 
\midrule
\textbf{Stellar Heterogeneity} & $T_{*, \, \rm{phot}}$, $T_{*, \, \rm{het}}$, $f_{\rm{het}}$ & $T_{*, \, \rm{phot}}$, $\Delta T_{*}$, $f_{\rm{het}}$ & $T_{*, \, \rm{phot}}$, $T_{*, \, \rm{het}}$, $f_{\rm{het}}$ \\[3pt]
\enddata 
\tablecomments{`H$^{-}$' denotes bound-free opacity of the hydrogen anion only. The free-free contribution arises when the H and e$^{-}$ abundances are included as separate free parameters.}
\tablerefs{\textbf{Line lists}: \citet{Polyansky2018}$^{1}$, \citet{Barber2006}$^{2}$, \citet{Li2015}$^{3}$, \citet{Tashkun2011}$^{4}$, \citet{Rothman2010}$^{5}$, \citet{Yurchenko2017}$^{6}$, \citet{Barber2014}$^{7}$, \citet{Yurchenko2011}$^{8}$, \citet{John1988}$^{9}$, \citet{Ryabchikova2015}$^{10}$, \citet{McKemmish2019}$^{11}$, \citet{McKemmish2016}$^{12}$, \citet{Wende2010}$^{13}$  }
\vspace{-10pt}
\end{deluxetable*}

Our retrievals include chemical species with prominent spectral features over the observed wavelength range \citep{Sharp2007,Tennyson2018} anticipated to be present in hot Jupiter atmospheres \citep{Madhusudhan2016}. For the near-infrared, we assess contributions from H$_2$O, CH$_4$, CO, CO$_2$, HCN, NH$_3$. For optical wavelengths, we consider Na, K, H$^{-}$, TiO, VO, FeH, and Fe. Each retrieval considered a subset of these potential species. Common opacity across all three codes are H$_2$O, H$^{-}$, collision-induced absorption due to H$_2$-H$_2$ and H$_2$-He \citep{Richard2012}, and H$_2$ Rayleigh scattering. 

Multiple studies have recently considered the inclusion of H$^{-}$ opacity in atmospheric retrievals \citep[e.g.,][]{Sotzen2020,Gandhi2020,Lothringer2020}. However, there remains no consensus on how to parametrize H$^{-}$ opacity in a retrieval context. H$^-$ is expected to become an important opacity source at high temperatures, when H$_2$ thermally dissociates to form atomic H \citep[e.g.,][]{Bell2018,Parmentier2018}. Atomic H absorbing a photon in the vicinity of a free electron produces free-free H$^{-}$ absorption \citep{Bell1987}:
\begin{equation}
    h\nu + \mathrm{e^-} + \mathrm{H} \rightarrow \mathrm{H} + \mathrm{e^-}
    \label{eqn:free-free}    
\end{equation}
Alternately, photodissociation of a bound H$^{-}$ ion results in bound-free H$^{-}$ absorption \citep{John1988}:
\begin{equation}
    h\nu + \mathrm{H^-} \rightarrow \mathrm{H} + \mathrm{e^-}
    \label{eqn:bound-free}    
\end{equation}
Although both processes are commonly referred to as `H$^{-}$ opacity', only the bound-free contribution involves a H$^{-}$ ion. Due to the distinct nature of these processes, we use a general treatment to parametrize H- opacity. Considering their combined opacity
\begin{equation}
    \kappa_{\rm{H}^{-}} = n_{\rm{H}^{-}} \, \sigma_{\rm{bf}, \, \rm{H}^{-}} + n_{\rm{H}} \, n_{\rm{e}^{-}} \, \alpha_{\rm{ff}, \, \rm{H}^{-}}
    \label{eqn:extinction_H-}  
\end{equation}
where $\kappa_{\rm{H}^{-}}$ is the H$^{-}$ extinction coefficient (in cm$^{-1}$), $\sigma_{\rm{bf}, \, \rm{H}^{-}}$ is the bound-free H$^{-}$ cross section (given in cm$^{2}$ by eqs. 4 and 5 from \citet{John1988}), $\alpha_{\rm{ff}, \, \rm{H}^{-}}$ is the free-free H$^{-}$ binary cross section (given in cm$^{5}$ by eq. 6 from \citet{John1988} multiplied by $k_{B} T$ in cgs units), and $n_{i}$ are the number densities (in cm$^{-3}$) of H$^{-}$, H, and free electrons. We propose that a parametrization suitable for free retrievals is to treat the mixing ratios of H$^{-}$, H, and e$^{-}$ as independent free parameters.

We also included the effects of unocculted spot/faculae in all our retrievals \citep[e.g.,][]{Rackham2018,Pinhas2018}. This was motivated by the negative slope towards blue wavelengths in our transmission spectrum (Figure~\ref{fig:transmission_spectrum}) - atmospheric scattering would instead cause a positive slope - and indicators of stellar activity for WASP-79 (Section~\ref{sec:stellar_activity}). We adopt a consistent prescription for stellar heterogeneity across all three codes. This invokes a three-parameter model, based on the approach of \citet{Pinhas2018}
\begin{equation}
    \Delta_{\lambda, \, \rm{obs}} = \Delta_{\lambda, \, \rm{atm}} \, \epsilon_{\lambda, \, \rm{het}}
    \label{eqn:stellar_contam_1}   
\end{equation}
where $\Delta_{\lambda, \, \rm{obs}}$ is the observed transmission spectrum, $\Delta_{\lambda, \, \rm{atm}}$ is the transmission spectrum from the planetary atmosphere alone, and $\epsilon_{\lambda, \, \rm{het}}$ is the wavelength-dependent `contamination factor' from a heterogeneous stellar surface. For a two-component stellar disc with a photosphere and an excess heterogeneity (spots or faculae), the contamination factor can be written as \citep{Rackham2018}
\begin{equation}
    \epsilon_{\lambda, \, \rm{het}} = \left(1 - f_{\rm het} \left(1 - \frac{I_{\lambda, \, \rm{het}} (T_{*, \, \rm{het}})}{I_{\lambda, \, \rm{phot}} (T_{*, \, \rm{phot}})} \right) \right)^{-1}
    \label{eqn:stellar_contam_2}   
\end{equation}
where $f_{\rm het}$ is the fractional stellar disc coverage of the heterogeneous regions, $I_{\lambda, \, \rm{het}}$ and $I_{\lambda, \, \rm{phot}}$ are the specific intensities of the heterogeneity and photosphere, respectively, with $T_{*, \, \rm{het}}$ and $T_{*, \, \rm{phot}}$ their corresponding temperatures. In our default retrieval prescription, we treat $f_{\rm het}$, $T_{*, \, \rm{het}}$, and $T_{*, \, \rm{phot}}$ as free parameters. We also investigated replacing the $T_{*, \, \rm{het}}$ parameter with the average temperature difference between heterogeneous regions and the photosphere: $\Delta T_{*} =  T_{*, \, \rm{het}} - T_{*, \, \rm{phot}}$. Since the stellar photosphere temperature is known \textit{a priori}, we place an informative Gaussian prior on $T_{*, \, \rm{phot}}$. We compute stellar spectra by interpolating models from the Castelli-Kurucz 2004 atlas \citep{Castelli2003} using the pysynphot package \citep{pysynphot}. 

All three codes conducted an atmospheric retrieval, as summarized in Table~\ref{tab:retrieval_config}, for parameter estimation. The full posterior distributions from these retrievals are provided as supplementary material. Detection significances for key model components were also computed, via Bayesian model comparisons. We now provide a brief summary of each retrieval code.

\subsubsection{NEMESIS}

The NEMESIS spectroscopic retrieval code \citep{Irwin2008} was originally developed for application to Solar System datasets. It combines a 1D parametrized radiative transfer model, using the correlated-k approximation \citep{Lacis1991}, with a choice of either optimal estimation \citep{Rodgers2000} or PyMultiNest \citep{Feroz2008,Feroz2009, Feroz2019,Buchner2014} for the retrieval algorithm \citep{Krissansen-Totton2018}. In this work we use the PyMultiNest version of NEMESIS.

The cloud parametrization used in NEMESIS follows that presented in \cite{Barstow2017} and \cite{Barstow2020}. The cloud is represented as a well-mixed slab constrained by top and bottom boundaries at variable pressures $P_{\rm top}$ and $P_{\rm base}$; the extinction efficiency is parametrized by a power law with a variable index, and the total optical depth is also retrieved. For the temperature profile, we use the parametrization presented in \cite{Guillot2010}. All other retrieved parameters are common to all three codes. The Gaussian prior for the stellar temperature for NEMESIS has a mean of 6600 K and a standard deviation of 500 K. 

\begin{figure*}[ht!]
    \centering
    \includegraphics[width=\textwidth, trim={0.0cm 0.2cm 0.0cm 0.1cm}]{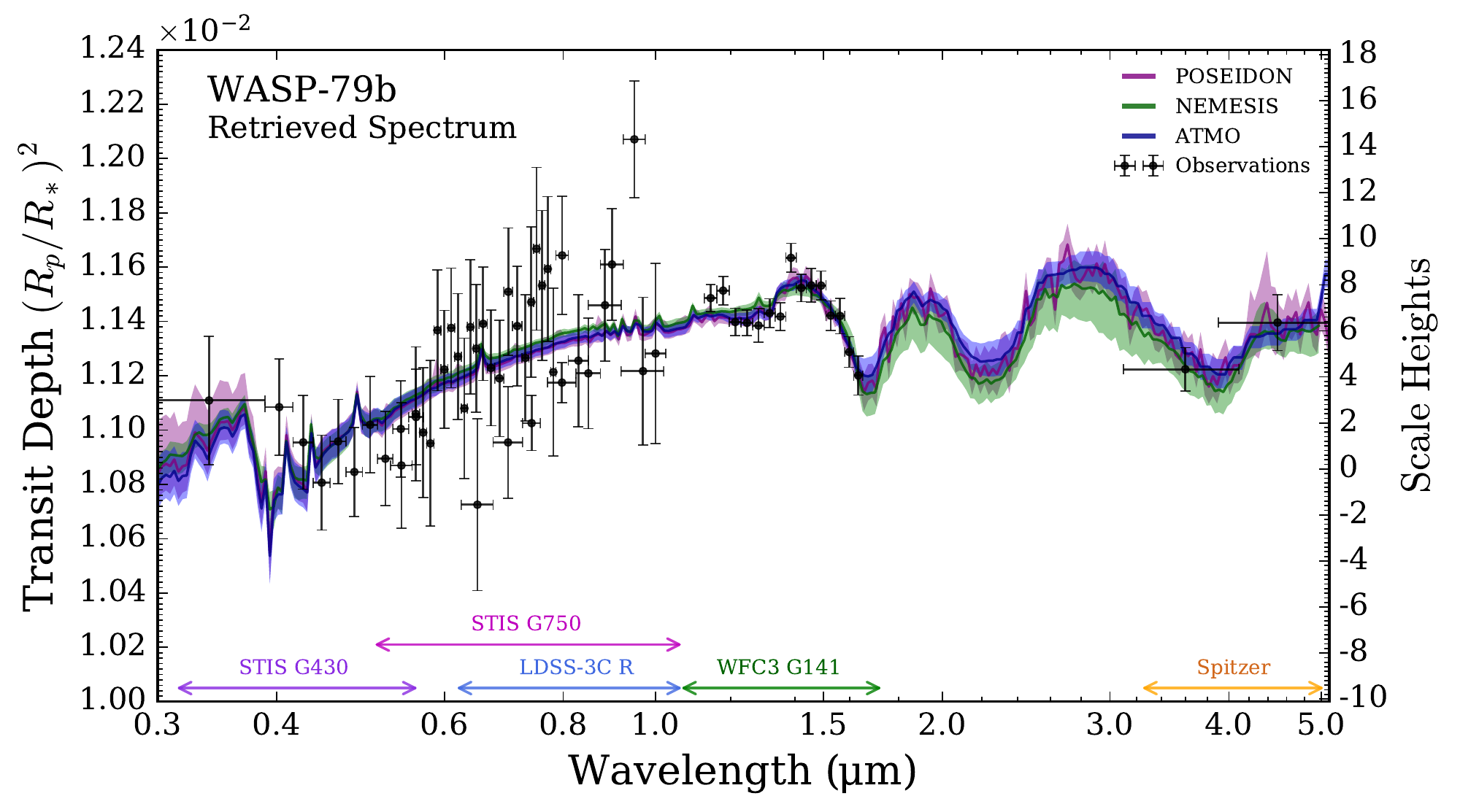}
    \caption{Atmospheric retrievals of WASP-79b's transmission spectrum. Retrieved model spectra are shown for three retrieval codes: POSEIDON (purple), NEMESIS (green), and ATMO (blue) \citep{MacDonald2017a,Irwin2008,Barstow2017,Amundsen2014,Wakeford2017}. The median retrieved spectra (solid lines) and 1$\sigma$ confidence regions (shading) from each code are binned to a common spectral resolution ($R = 100$). The spectral range for each instrument mode comprising the observations are indicated at the base of the plot. The preferred interpretation consists of H$_2$O opacity in the infrared, with the combination of H$^{-}$ opacity and the influence of unocculted stellar faculae in the visible.} 
    \label{fig:retrievals}
\end{figure*}

\subsubsection{POSEIDON}

POSEIDON \citep{MacDonald2017a} is a radiative transfer and retrieval code designed to invert exoplanet transmission spectra. Applications in the literature range from hot Jupiters to terrestrial planets \citep[e.g.,][]{Kilpatrick2018,MacDonald2019,Kaltenegger2020}. Radiative transfer is computed via the sampling of high spectral resolution ($R \sim 10^6$) cross sections. Over 50 chemical species are supported as retrievable parameters, of which 15 are employed in this study. The P-T profile is parametrized as in \citet{Madhusudhan2009}. Clouds and hazes are parametrized according to the inhomogenous cloud prescription in \citet{MacDonald2017a}. The stellar photosphere temperature, $T_{*, \, \rm{phot}}$, has a Gaussian prior with a 6600\,K mean and 100\,K standard deviation. The stellar heterogeneity temperature, $T_{*, \, \rm{het}}$, has a uniform prior from 60-140\% of the \emph{a priori} mean photosphere temperature. The heterogeneity coverage fraction, $f_{\rm het}$, has a uniform prior from $0.0 - 0.5$. The 30-dimensional parameter space is explored using the nested sampling algorithm MultiNest \citep{Feroz2008,Feroz2009,Feroz2019}, as implemented by PyMultiNest \citep{Buchner2014}. 

\subsubsection{ATMO}

The ATMO forward model \citep{Tremblin2015,Drummond2016,Goyal2018} has been used previously as a spectroscopic retrieval model for both transmission and emission spectra \citep[e.g][]{Wakeford2017,Evans2017}. We assume isothermal temperature profiles and perform free chemistry retrieval in this work. The cloud parametrization and opacities used in ATMO for our retrieval are described in \citet{Goyal2018,Goyal2019}. The stellar heterogeneity parameter priors are the same as described for POSEIDON above. In previous works, ATMO has employed the MCMC retrieval algorithm within EXOFAST \citep{Eastman2013}. Here, we have updated ATMO to use the nested sampling code dynesty \citep{Speagle2020}.

\subsection{Retrieval Results}

\begin{figure*}[ht!]
    \centering
    \includegraphics[width=\textwidth, trim={0.0cm 0.2cm 0.0cm 0.1cm}]{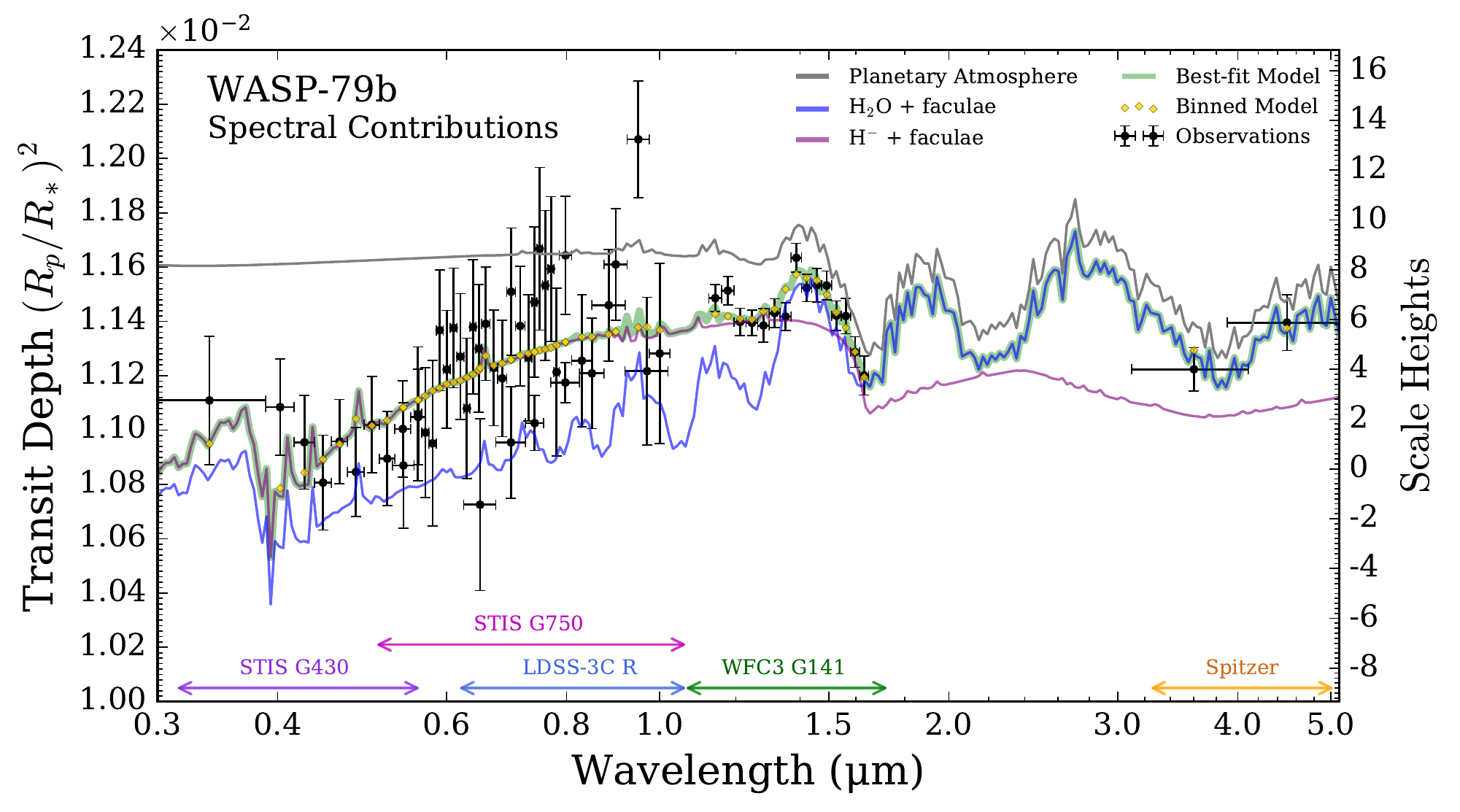}
    \caption{Contributions to the best-fitting model of the transmission spectrum of WASP-79b. The maximum likelihood retrieved spectrum (green shading) is decomposed into the following sub-models: (i) the planetary atmosphere spectrum, without contributions from faculae (grey); (ii) H$_2$O opacity and faculae, without contributions from H$^{-}$ opacity (blue); and (iii) H$^{-}$ opacity and faculae, without contributions from H$_2$O opacity (purple). Note that all four models include H$_2$-H$_2$ collision-induced absorption (CIA), seen most clearly for the `H$^{-}$ + faculae' model redwards of 1.64\,\micron. The best-fitting model, binned to the resolution of the observations, is overlaid for comparison (gold diamonds). The spectra come from the `minimal' POSEIDON model (see text for details) for illustration purposes. The best-fitting solutions from NEMESIS and ATMO are similar.}
    \label{fig:spectral_contributions}
\end{figure*}

\begin{figure*}[ht!]
    \includegraphics[width=0.98\textwidth]{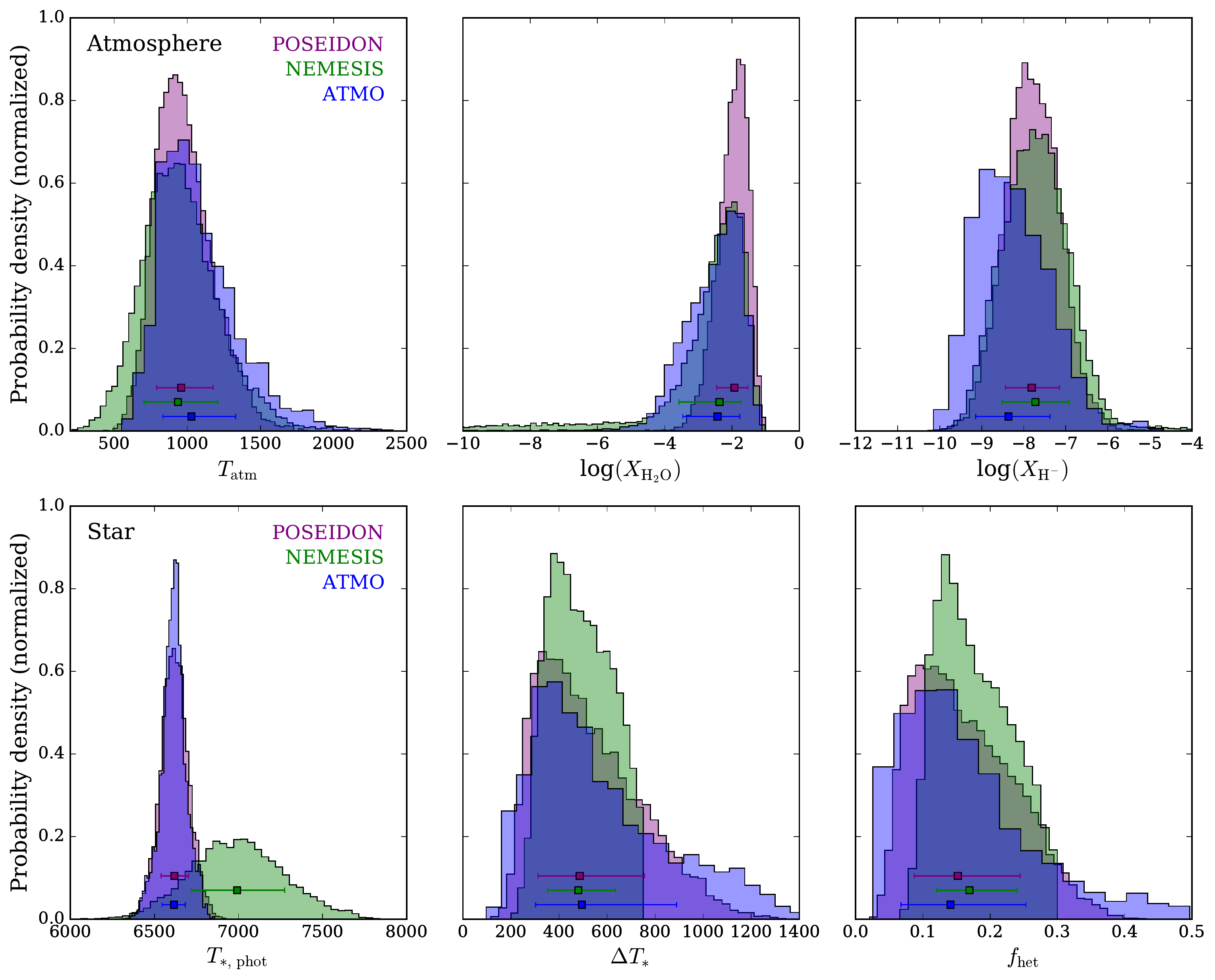}
    \caption{Retrieved model parameters from the transmission spectrum of WASP-79b. Posterior distributions from three retrieval codes are compared: POSEIDON (purple), NEMESIS (green), and ATMO (blue). Top panels: retrieved planetary atmosphere properties. $T_{\rm{atm}}$ represents either an isothermal temperature (ATMO) or the 1\,mbar temperature where a parametrized temperature profile is used (NEMESIS and POSEIDON). $X_{\rm{i}}$ are the volume mixing ratios of H$_2$O and H$^{-}$, respectively, plotted on a log$_{10}$ scale. Bottom panels: retrieved stellar properties. $T_{*, \, \rm{phot}}$, $\Delta T_{*}$, and $f_{\rm{het}}$ are the stellar photosphere temperature, heterogeneity-photosphere temperature difference, and heterogeneity coverage fraction, respectively. For each parameter, the median retrieved value (squares) and $\pm 1\,\sigma$ confidence regions (error bars) are overlaid.} 
    \label{fig:posteriors}
\end{figure*}

\begin{deluxetable*}{lcccc} \label{tab:retrieval_results_summary}
    \tablecaption{Retrieved Planetary and Stellar Parameters}
    \tablewidth{0pt}
    \tablehead{
    Retrieval & POSEIDON & NEMESIS & ATMO & Minimal
    }
    \startdata \\[-8pt]
    \textbf{Planetary Atmosphere} \\ 
    \hspace{0.5em} $T_{\rm{1 \, mbar}}$ (K) & $958^{+217}_{-168}$ & $936^{+272}_{-233}$ & $1028^{+303}_{-196}$ & $836^{+187}_{-139}$ \\
    \hspace{0.5em} $R_{\rm{p, \, ref}}$ ($R_J$) & $1.68^{+0.02}_{-0.02}$ & $1.62^{+0.02}_{-0.03}$ & $1.75^{+0.01}_{-0.02}$ & $1.70^{+0.01}_{-0.02}$ \\
    \hspace{0.5em} log($X_{\rm{H_2 O}}$) & \hspace{-1.0em} $-1.92^{+0.39}_{-0.53}$ & \hspace{-1.0em} $-2.37^{+0.64}_{-1.19}$ & \hspace{-1.0em} $-2.43^{+0.66}_{-1.03}$ & $-2.43^{+0.66}_{-0.96}$ \\
    \hspace{0.5em} log($X_{\rm{H^{-}}}$) & \hspace{-1.0em} $-7.80^{+0.66}_{-0.63}$ & \hspace{-1.0em} $-7.72^{+0.78}_{-0.79}$ & \hspace{-1.0em} $-8.37^{+0.98}_{-0.78}$ & $-8.81^{+0.78}_{-0.65}$ \\[3pt]
    \midrule
    \textbf{Derived Properties} \\
    \hspace{0.5em} O/H ($\times$ stellar) & $13.9^{+19.9}_{-9.7}$ & $4.7^{+16.0}_{-4.4}$ & $4.1^{+14.8}_{-3.8}$ & $4.1^{+14.8}_{-3.7}$ \\[3pt]
    \midrule
    \textbf{Stellar Properties} \\ 
    \hspace{0.5em} $T_{*, \, \rm{phot}}$ (K) & \hspace{-1.2em} $6619^{+84}_{-80}$ & \hspace{-0.8em} $6992^{+281}_{-271}$ & \hspace{-0.8em} $6616^{+67}_{-71}$ & $6623^{+87}_{-92}$ \\
    \hspace{0.5em} $\Delta T_{*}$ (K) & $486^{+270}_{-173}$ & $480^{+153}_{-128}$ & $495^{+390}_{-192}$ & $452^{+605}_{-237}$ \\
    \hspace{0.5em} $f_{\rm{het}}$ & $0.15^{+0.09}_{-0.07}$ & $0.17^{+0.07}_{-0.05}$ & $0.14^{+0.11}_{-0.07}$ & $0.15^{+0.20}_{-0.10}$ \\[3pt]
    \midrule
    \textbf{Statistics} \\ 
    \hspace{0.5em} $\chi^2_{\nu, \, \rm{min}}$ & $1.84$ & $1.68$ & $1.25$ & $1.12$ \\
    \hspace{0.5em} $N_{\rm param}$ & $30$ & $21$ & $9$ & $7$ \\
    \hspace{0.5em} Degrees of freedom & $31$ & $40$ & $52$ & $54$ \\[3pt]
    \enddata 
    \tablecomments{ Only parameters with bounded constraints (i.e., both lower and upper bounds) are included in this summary table - see the online supplementary material for full posterior distributions. The `minimal' retrieval is the simplest model that can fit our transmission spectra of WASP-79b: a clear, isothermal, atmosphere containing H$_2$O and H$^{-}$ alongside stellar contamination from unocculted faculae (7 free parameters) - also computed with POSEIDON. $R_{\rm{p, \, ref}}$ is defined at $P =$ 10 bar for NEMESIS and POSEIDON, and 1 mbar for ATMO. $\Delta T_{*} =  T_{*, \, \rm{het}} - T_{*, \, \rm{phot}}$. The stellar O/H is assumed equal to WASP-79's stellar [Fe/H] (0.03, \citet{Stassun2017}).}
    \vspace{-10pt}
\end{deluxetable*}

Here we present the results of our comparative retrievals. We first explore the best-fitting atmospheric and stellar interpretation matching the transmission spectrum of WASP-79b. Constraints on the atmospheric properties of WASP-79b are then presented, followed by inferences of stellar heterogeneity.  

\subsubsection{Explaining the Transmission Spectrum of WASP-79b}

Our retrievals arrived at a consistent explanation for the transmission spectrum of WASP-79b. Our best-fitting model spectra, shown in Figure~\ref{fig:retrievals}, are characterized by three components: (i) H$_2$O opacity (explaining the absorption feature around $1.4~\micron$); (ii) spectral contamination from unocculted faculae (producing the negative slope over optical wavelengths); and (iii) H$^{-}$ bound-free absorption (resulting in a relatively smooth continuum from 0.4 - 1.3 $\micron$). Our new STIS observations play a crucial role in the identification of faculae, extending the coverage of WASP-79b's transmission spectrum to wavelengths $<~0.6~\micron$ where the effects of stellar contamination are more pronounced. We verified that retrievals excluding the LDSS3 data (i.e., STIS + WFC3 + Spitzer only) arrive at the same conclusion.

The broad agreement between our retrievals, despite their quite different configurations (Table~\ref{tab:retrieval_config}), motivated an exercise to identify the minimal model capable of explaining the present observations. Although the fit qualities shown in Figure~\ref{fig:retrievals} are comparable, the differing numbers of free parameters (30 for POSEIDON, 21 for NEMESIS, and 9 for ATMO) resulted in a range of best-fitting reduced chi-square values suggestive of model over-complexity for the present datasets ($\chi^2_{\nu, \, \rm{min}}$ = 1.84, 1.68, and 1.25 for POSEIDON, NEMESIS, and ATMO, respectively). Taking the 9-parameters defining the ATMO model as a starting point, we ran additional retrievals with progressively fewer free parameters to identify the simplest model capable of explaining WASP-79b's transmission spectrum - corresponding to the model with maximal Bayesian evidence (analogous to $\chi^2_{\nu, \, \rm{min}}$ minimization). This process arrived at a 7 parameter `minimal' model\footnote{POSEIDON computed the `minimal' retrieval, but the similar NEMESIS and ATMO fits would lead to the same conclusion.}: H$_2$O and H$^{-}$ in a clear, isothermal, H$_2$-dominated atmosphere transiting a stellar surface with unocculted faculae ($\chi^2_{r, \, \rm{min}}$ = 1.12). We show the best-fitting spectrum from this retrieval in Figure~\ref{fig:spectral_contributions}, demonstrating consistency with the more complex models explored previously. With respect to this minimal model, Bayesian model comparisons yielded strong detections of faculae ($4.7\sigma$) and H$_2$O ($4.0\sigma$), along with moderate evidence of H$^{-}$ ($3.3\sigma$).

The contributions of these opacity / contamination sources to the best-fitting minimal model spectrum are shown in Figure~\ref{fig:spectral_contributions}. The features of the observed spectrum are reproduced by a combination of H$_2$O, H$^{-}$ and H$_2$-H$_2$ collision-induced absorption within the planet's atmosphere, alongside contributions from unocculted faculae on the stellar surface. Faculae occupying regions of the star outside the transit chord result in the planet occulting a region of the star that is cooler and darker than the disc average, since faculae are relatively hot and bright. This results in an underestimation of the true planet-to-star radius ratio, as illustrated by the `atmosphere only' model in Figure~\ref{fig:spectral_contributions}, which has greater transit depths than the composite spectrum. The magnitude of this effect varies with wavelength, such that the faculae/disc contrast is most pronounced at shorter wavelengths, resulting in shallower transit depths in the near-UV and optical relative to the infrared, thus giving rise to the negative slope at short wavelengths. This `transit light source effect' is discussed more generally in \cite{Rackham2018}. 

\subsubsection{The Atmosphere of WASP-79b}

Our retrieved atmospheric properties\footnote{Full posteriors are available in the supplementary material.} are displayed in Figure~\ref{fig:posteriors} (top row) and summarized in Table~\ref{tab:retrieval_results_summary}. All three codes reach excellent agreement on the atmospheric parameters of WASP-79b, which all agree within their respective $1\sigma$ confidence intervals.

Our robust detection of H$_2$O allows the atmospheric metallicity of WASP-79b to be constrained. We derive statistical constraints on the atmospheric O/H ratio from the full set of posterior samples as in \citet{MacDonald2019}. Taking the stellar [Fe/H] \citep[= 0.03,][]{Stassun2017} to be representative of the stellar [O/H], we compute the metallicity of WASP-79b (relative to its host star) via M = $\mathrm{(O/H)_{atm} / (O/H)_{*}}$. Our retrieved metallicities are then as follows: $13.9^{+19.9}_{-9.7} \times$ stellar (POSEIDON), $4.7^{+16.0}_{-4.4} \times$ stellar (NEMESIS), and $4.1^{+14.8}_{-3.8} \times$ stellar (ATMO). The median retrieved values are suggestive of a super-stellar metallicity for WASP-79b. However, a stellar metallicity remains consistent with the present observations to $1\sigma$ for NEMESIS and ATMO (due to the long tails in their H$_2$O abundance posteriors), and to $2\sigma$ for POSEIDON. 

The bound-free absorption of H$^{-}$ inferred from our retrievals produces corresponding constraints on its abundance. All three retrievals concur on a H$^{-}$ abundance of log($X_{\rm{H^{-}}}$) $\approx -8.0 \pm 0.7$ (see Table~\ref{tab:retrieval_results_summary}). This precise H$^{-}$ constraint arises from two principal features of our observations: (i) the high-precision WFC3 G141 data ($\sim$ 50 ppm) closely follows the shape of the H$^{-}$ bound-free opacity near the photodissociation limit (see Figure~\ref{fig:spectral_contributions}); and (ii) the long spectral baseline provided by our STIS observations, alleviating normalization degeneracies \citep[see, e.g.,][]{Heng2017,Welbanks2019a}. We discuss the plausibility of our inferred H$^{-}$ opacity in Section~\ref{subsec:discussion_H_minus}.

Our retrievals additionally constrain the terminator temperature of WASP-79b. Despite the three different P-T profile prescriptions (see Table~\ref{tab:retrieval_config}), all three codes arrived at the same conclusion: a near-isothermal terminator with $T \sim 1000 \pm 300$\,K. We summarize the retrieved temperatures for each code in Table~\ref{tab:retrieval_results_summary}, where for the non-isothermal profiles we quote $T_{\rm{1 \, mbar}}$ as a photosphere proxy. This retrieved temperature is markedly colder than the equilibrium temperature of WASP-79b ($1900~\pm~50~\text{K}$, \citealt{Smalley2012}). Recently, \citet{MacDonald2020} noted that most retrieved temperatures from transmission spectra are significantly colder than $T_{\rm eq}$. They attributed this trend to a bias arising from 1D atmosphere assumptions. We discuss the impact of this bias on our retrieved metallicity in Section~\ref{subsec:discussion_metallicity}.

The atmospheric region probed by our transmission spectrum is consistent with a lack of detectable cloud opacity. However, the limits on cloud properties derived by our retrievals (e.g., $P_{\rm cloud} > 10^{-5}$\,bar to $2\sigma$ from POSEIDON) still allow for the existence of deeper cloud decks. Nevertheless, for the present observations, our results are invariant to the chosen cloud prescription.

\newpage

\subsubsection{Stellar Properties}

Our retrieved stellar properties are shown in Figure~\ref{fig:posteriors} (bottom row) and also summarized in Table~\ref{tab:retrieval_results_summary}. All three retrievals produce a consistent interpretation: a stellar surface with $\sim 15\%$ faculae coverage at a temperature contrast of $\sim 500$\,K. The only disagreement found is the retrieved photospheric temperature, for which NEMESIS retrieves a value $\sim 400$\,K higher than ATMO and POSEIDON. This difference arises from the Gaussian prior used for $T_{*, \, \rm{phot}}$ by POSEIDON and ATMO having a standard deviation $1/5^{\rm{th}}$ that used by NEMESIS (100\,K vs. 500\,K). However, this discrepancy does not influence the agreement for any of the other retrieved parameters, as it is the temperature contrast, $\Delta T_{*}$, which governs the manifestation of spectral contamination from heterogeneous regions. 

\begin{figure}[ht!]
    \centering
    \includegraphics[width=\columnwidth]{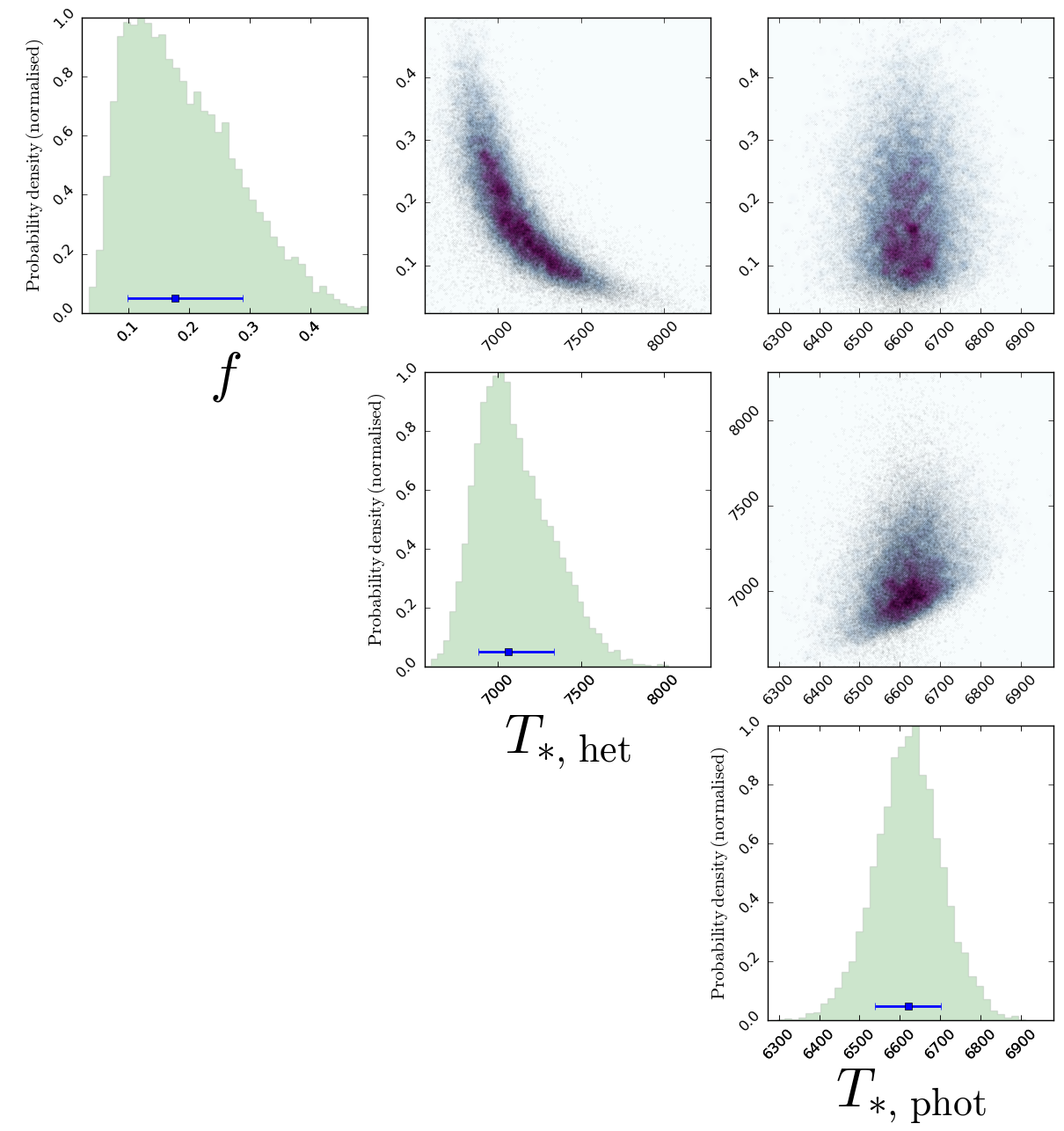}
    \caption{Correlations between retrieved stellar parameters. The corner plot shows a subset of the posterior distribution from the `full' POSEIDON retrieval model (all full posteriors are available in the supplementary material). The histograms correspond to the same (purple) histograms shown in Figure~\ref{fig:posteriors}. The most significant correlation is a curved degeneracy between the faculae coverage fraction and the faculae temperature - indicating hotter faculae occupying a lower area produce similar quality spectral fits to cooler faculae occupying a greater area.}
    \label{fig:stellar_correlations}
\end{figure}

The degeneracies between our retrieved stellar parameters are shown in Figure~\ref{fig:stellar_correlations}. The fractional faculae coverage, $f_{\rm{het}}$, and faculae temperature, $T_{*, \, \rm{het}}$, are partially degenerate. The origin of this degeneracy is intuitive: a large coverage fraction with relatively cool faculae produces a similar contamination signal to a lower coverage fraction with hot faculae. The uncertainties in other parameters introduced by this degeneracy are already accounted for in the marginalized posteriors shown in Figure~\ref{fig:posteriors}. The $f_{\rm{het}}$ - $T_{*, \, \rm{het}}$ degeneracy is not, however, an exact degeneracy. For sufficiently large $\Delta T_{*}$, wavelength-dependent spectral signatures arise (from the intensity ratio in Equation~\ref{eqn:stellar_contam_2}) that cannot be compensated by varying $f_{\rm{het}}$ \citep[see e.g.,][]{Pinhas2018}. Such signatures are especially prominent at the short wavelengths sampled by our STIS G430L observations (0.3 - 0.6 $\micron$, see Figure~\ref{fig:retrievals}). This second-order effect, crucially probed by STIS, allows bounded constraints on the faculae coverage fraction.

\section{Discussion} \label{discussion}

\subsection{Comparison with \citet{Sotzen2020} \\ and \citet{Skaf2020}} \label{subsec:discussion_literature_comparison}

An optical to infrared transmission spectrum of WASP-79b, combining results from Hubble/WFC3 with optical data from LDSS3 and photometry from TESS and Spitzer, has been published by \cite{Sotzen2020}. A retrieval analysis on this dataset was performed using the ATMO retrieval code. Our present study extends the blue wavelength coverage of the transmission spectrum of WASP-79b from 0.6 $\micron$ down to 0.3 $\micron$ with Hubble/STIS G430L data while adding STIS G750L data to complement the prior LDSS3 observations.

The ATMO free retrieval presented by \cite{Sotzen2020} considered many similar atmospheric parameters to those used here, including H$_2$O, CO, Na, K, VO, FeH, and H$^{-}$. Our retrievals investigated a wider range of chemical species, cloud, and temperature profile parametrizations, and a stellar heterogeneity treatment (see Table~\ref{tab:retrieval_config}). Despite our different retrieval prescriptions and expanded dataset, we obtain consistent H$_2$O abundances, temperatures, and agree on the overall lack of significant cloud opacity. However, differences emerge when considering other gaseous species. \cite{Sotzen2020} infer the presence of Na, although they stress that this is driven by the TESS photometry point having a deeper transit than the LDSS3 data, rather than a resolved Na profile. We do not find evidence for Na. The other key difference is that \cite{Sotzen2020} do not detect H$^{-}$, but instead find evidence for FeH. 

These differences can be attributed to the additional information provided by our STIS observations. The combination of the WFC3 spectrum with the broad, relatively flat, STIS spectra leads our retrievals to favor H$^{-}$ as the optical absorber (modulated by unocculted faculae), rather than FeH. Whilst both FeH and H$^{-}$ are capable of fitting the short wavelength end of the WFC3 spectrum, FeH absorption does not extend shortwards of $\sim 0.7 \micron$ (see Figure 14 in \citealt{Sotzen2020} and \citealt{Tennyson2018}). The importance of H$^{-}$ is seen clearly in Figure~\ref{fig:spectral_contributions}, and is identified by all three retrieval codes, as shown in Figure~\ref{fig:retrievals}. We performed an additional POSEIDON retrieval without H$^{-}$ opacity to investigate the differences between our results and \citet{Sotzen2020}. In this case, FeH is recovered with a large abundance mode ($\log(\rm{FeH}) = -2.8 \pm 0.5$) consistent with that found by \cite{Sotzen2020} (see the supplementary material). However, our retrievals considering both H$^{-}$ and FeH rule out such high FeH abundances ($\log(\rm{FeH}) < -3.78$ to $2\sigma$) and have a higher Bayesian evidence. We also note that FeH abundances exceeding $\sim 10^{-7}$ are unexpected in thermochemical equilibrium for a giant planet at WASP-79b's equilibrium temperature \citep{Visscher2010}, and FeH abundances $> 10^{-6}$ were recently ruled out by high-resolution transmission spectra of 12 giant exoplanets \citep{Kesseli2020}. In summary, we find that the combined effect of H$^{-}$ and unocculted faculae provides a more robust explanation of WASP-79b's transmission spectrum. 

Recently, \citet{Skaf2020} retrieved an alternative reduction of WASP-79b's WFC3 transmission spectrum with the TauREx code. They also do not include H$^{-}$ opacity, and hence recover a similar FeH abundance to \cite{Sotzen2020}. The retrieved temperature and H$_2$O abundance are consistent with our findings.

\subsection{Plausibility of H$^{-}$} \label{subsec:discussion_H_minus}

H$^{-}$ is thought to become an important opacity source for high temperature planets when H$_2$ thermally dissociates to form atomic hydrogen. H$_2$ dissociation is generally expected to occur on the daysides of `ultra-hot' Jupiters (e.g., \citealt{Bell2018}), with prominent H$^{-}$ opacity expected around 2500~K \citep{Arcangeli2018}. Under the assumption of chemical equilibrium, H$^{-}$ abundances decrease for lower temperatures, with observable signatures in transmission spectra not expected below $\sim$ 2100~K \citep{Goyal2020}. 

Our retrieved terminator temperature for WASP-79b ($\sim$ 1000 K) lies in a regime where H$^{-}$ opacity would not be expected under equilibrium considerations. In comparison, the self-consistent models of \citet{Goyal2020} predict 1 mbar temperatures of $\sim$ 2000~K for WASP-79b (assuming C/O = 0.5, M/H = 10 $\times$ solar, and recirculation factor of unity). However, our three retrieval analyses assumed a uniform (1D) composition and temperature across the terminator. Transmission spectra of planets exhibiting non-uniform compositions in the terminator region can lead to biased temperatures when subject to a 1D atmospheric retrieval \citep{MacDonald2020,Pluriel2020}. Given the strong temperature dependence of equilibrium H$^{-}$ abundances, one would expect large H$^{-}$ compositional gradients between different regions of the terminator. \citet{MacDonald2020} demonstrated that transmission spectra of planets for which H$^{-}$ is only present on the warmer evening terminator result in retrieved 1D temperatures biased $\sim$ 1000~K colder than the terminator average temperature. An inter-terminator H$^{-}$ abundance gradient therefore provides a potential explanation for the discrepancy between our retrieved temperatures, self-consistent models, and the existence of H$^{-}$ opacity. 

A further possibility is that the $\sim 10^{-8}$ H$^{-}$ abundance we infer results from disequilibrium photochemistry. \citet{Lewis2020} recently showed that a similar H$^{-}$ abundance can be produced in HAT-P-41b's atmosphere ($T_{\rm eq} \sim 1900$\,K) by photochemical production of free electrons followed by dissociative electron attachment of H$_2$ ($\mathrm{H_2+e^-\rightarrow H +H^-}$).

\subsection{The Metallicity of WASP-79b in Context} \label{subsec:discussion_metallicity}

An essential goal of atmospheric studies is to relate retrieved properties to planetary formation histories and environments. The abundance enhancements of elements relative to hydrogen, or metallicity, provides a crucial link to planetary formation mechanisms  \citep[e.g.,][]{Oberg2011,Madhusudhan2014,Mordasini2016}. The Solar System giant planets exhibit an inverse correlation between planet mass and metallicity (from C/H measurements), commonly interpreted as evidence of formation by core-accretion \citep{Pollack1996}. On the other hand, many hot Jupiter exoplanets are consistent with sub-stellar O/H ratios below the Solar System mass-metallicity trend \citep{Barstow2017,Pinhas2018,Welbanks2019b}. With a roughly Jovian mass (0.9 $\text{M}_\text{J}$), WASP-79b provides an opportunity to benchmark a hot Jupiter metallicity against elemental abundance measurements of Jupiter.   

Our retrieved H$_2$O abundances generally suggest somewhat super-stellar O/H ratios for the atmosphere of WASP-79b ($\sim 0.3 - 34 \times$ stellar). This is consistent with the C/H abundance of Jupiter of $\sim 4 \times$ solar \citep{Atreya2018} and recent preliminary measurements of the equatorial O/H abundance of Jupiter from JUNO \citep{Li2020}. However, our median H$_2$O abundances are $\sim 100 \times$ higher than those derived from transmission spectra of the similar mass hot Jupiters HD~209458b and HD~189733b \citep{Barstow2017,MacDonald2017a,Pinhas2018,Welbanks2019b}. We note that our retrieved H$_2$O abundances, and hence metallicities, may be biased by $\lesssim$ 1 dex towards higher abundances if a H$^{-}$ abundance gradient exists between the morning and evening terminators \citep[see][their Figure 3]{MacDonald2020}. Even accounting for a factor of 10 H$_2$O bias, the maximum likelihood H$_2$O abundance for WASP-79b would remain $\sim 10 \times$ higher than those of HD~209458b and HD~189733b. This suggests the formation of WASP-79b may be more analogous to Jupiter than to other similar mass hot Jupiters, indicative of a diversity of formation avenues at play across the hot Jupiter population.

\subsection{Faculae Characteristics} \label{subsec:discussion_faculae}

Our retrievals favor models including stellar contamination, arising from unocculted faculae $\approx$ 500\,K hotter than the photosphere of the host star and covering $\approx$ 15\% of the stellar surface (Table~\ref{tab:retrieval_results_summary}). A similar contamination effect was also observed in the transmission spectrum of GJ~1214b \citep{Rackham2017}, though that study found $\Delta T \approx$ 350\,K and a faculae coverage fraction about five times smaller ($\approx$ 3\%). However, it is hard to reconcile the presence of faculae on the surface of WASP-79 with other available observations: although our XMM-Newton observations indicate a moderate level of chromospheric activity, the stellar photosphere is not expected to present large active regions -- as indicated by the low photometric variations in the TESS light curves described in Section~\ref{sec:stellar_activity}. However, certain geometrical configurations of the system can resolve this apparent discrepancy. As WASP-79b follows a nearly polar orbit \citep{Addison2013}, we do not know the inclination of WASP-79. This opens the possibility that we are observing the star pole-on. We do not expect to see high levels of variability in the TESS data in this case, as effectively few active regions would rotate into or out of view. This scenario would explain the low-level photometric variability, while still allowing a high coverage fraction of unocculted faculae. While this remains speculative, we also note that the posteriors for our retrieved stellar parameters are broad. Consequently, a lesser degree of heterogeneity is still consistent with our transmission spectrum (e.g. $\Delta T \approx$ 300\,K temperature contrast with $\approx$ 7\% coverage fraction lies within $1\,\sigma$). A promising avenue for future work would be to better constrain the degree of stellar heterogeneity WASP-79 exhibits via out-of-transit stellar decomposition (e.g., \citealt{Wakeford2019,Iyer2020}).

\subsection{Prospects for the potential JWST ERS Observations} \label{subsec:discussion_JWST}

WASP-79b is a shortlisted target for the JWST Transiting Exoplanet Community ERS Program (PI: Batalha, ERS 1366, \citealt{Bean2018}). If observed, a complete transmission spectrum will be constructed from 0.6 - 5.3 $\micron$ via observations with four instrument modes: NIRISS SOSS, NIRSpec G235H, NIRSpec G395H, and NIRCam F322W2. Our results inform the potential science return of such observations.

Consistent with previous studies \citep{Sotzen2020,Skaf2020}, we find an atmosphere with a high H$_2$O abundance ($\sim 1\%$) and negligible cloud opacity. Our best-fitting models, therefore, predict a prominent 3~$\micron$ H$_2$O feature spanning $\sim$ 4 scale heights (see Figure~\ref{fig:spectral_contributions}), which will be readily detectable by all four JWST observations. Consequently, precise H$_2$O abundance and metallicity determinations ($\lesssim$ 0.2 dex, \citealt{Sotzen2020}) will be possible. Spectrally-resolved CO and CO$_2$ features around 4.5 $\micron$ with NIRSpec G395H will further allow a precise C/O ratio determination.   

Our tentative inference of H$^{-}$ opacity offers intriguing possibilities for the potential JWST observations of WASP-79b. First, NIRISS SOSS can readily assess the presence of H$^{-}$ via precision measurements of the characteristic bound-free opacity in the optical and near-infrared. If confirmed, a high-significance H$^{-}$ detection and abundance constraint would result. Secondly, the longest wavelength observations ($>~4~\micron$) with NIRSpec G395H may detect free-free H$^{-}$ opacity, enabling one to measure the atmospheric electron mixing ratio \citep{Lothringer2020}. JWST observations of WASP-79b, therefore, have the potential to open a window into ionic chemistry in hot Jupiter atmospheres.

\section{Summary} \label{summary}

We presented a new optical transmission spectrum of the hot Jupiter WASP-79b using data from three HST/STIS transits obtained with the G430L and G750L gratings. We introduced a new data-driven Bayesian model comparison approach to optimize Gaussian process kernel selection and applied it to correct for systematics in our light curve data analysis. We combined our observations with LDSS3, HST/WFC3, and Spitzer data from \citet{Sotzen2020} to yield a complete transmission spectrum from the near-UV to infrared ($0.3 - 5 \micron$). We subjected this spectrum to a series of atmospheric retrievals with three different codes to infer properties of the host star and the planetary atmosphere. Our main findings are as follows: 

\begin{itemize}
    \item Our measured HST/STIS transmission spectrum shows a peculiar slope: transit depths decrease towards blue wavelengths throughout the optical. A similar slope was observed by \citet{Sotzen2020} using ground-based LDSS3 data, with our observations extending the range down to 0.3 $\micron$.

    \item XMM-Newton/OM UV observations of WASP-79 suggests some UV stellar activity, suggesting the presence of spots/faculae in the stellar chromosphere. We therefore included a simple model describing the chromatic effects that unocculted spots/faculae would have on the measured transmission spectrum within our retrievals. Our best-fitting model prefers a solution with $\sim$ 15\% faculae coverage $\sim 500$\,K hotter than the stellar photosphere. Though auxiliary optical-wavelength photometric observations indicate low-level stellar variability, this may be consistent with our inferred heterogeneity if WASP-79 has a near pole-on viewing geometry.

    \item Our retrievals all find a near-isothermal terminator with $T \sim 1000 \pm 300$\,K, a somewhat super-stellar metallicity, and that WASP-79b's atmosphere is best described by a combination of H$_2$O and H$^{-}$. Our retrievals infer a H$_2$O abundance of $\sim$ 1\% - in agreement with previous studies - and a H$^{-}$ abundance of log($X_{\rm{H^{-}}}$) $\approx -8.0 \pm 0.7$. Our inclusion of HST/STIS data causes the retrievals to prefer H$^{-}$ and unocculted faculae over the previously suggested FeH opacities. 

    \item WASP-79b is one of the shortlisted targets for the JWST Transiting Exoplanet Community ERS program. We predict a H$_2$O feature of $\sim$ 4 scale heights at 3~$\micron$ would be accessible to near-infrared JWST observations. Furthermore, abundance determinations of CO and CO$_2$ around 4.5 $\micron$ would allow for precise C/O ratio determinations, which consequently could be linked to the formation and migration history of WASP-79b. Finally, our inference of H$^{-}$ offers the intriguing possibility that JWST transmission spectra can directly measure the abundances of ionic species in hot Jupiter atmospheres.

\end{itemize}

\software{george \citep{Ambikasaran2015}, MultiNest \citep{Feroz2008,Feroz2009,Feroz2019}, PyMultiNest \citep{Buchner2014}, BATMAN \citep{Kreidberg2015}, Limb Darkening Toolkit (LDTk) \citep{Parviainen2015}, pysynphot \citep{pysynphot}, NEMESIS \citep{Irwin2008}, POSEIDON \citep{MacDonald2017a}, ATMO \citep{Tremblin2015,Drummond2016,Goyal2018}}, Astropy \citep{2013A&A...558A..33A,2018AJ....156..123A}, ISIS \citep{Houck2000}, XMM-Newton Science Analysis System.\footnote{\href{https://xmm-tools.cosmos.esa.int/external/xmm\_user\_support/documentation/sas\_usg/USG/}{XMM-Newton SAS: User Guide}}

\section*{Acknowledgements}
We extend gratitude to the anonymous referee for a constructive report that improved our study. G. W. H. acknowledges long-term support from NASA, NSF, Tennessee State University, and the State of Tennessee through its Centers of Excellence Program. JSF acknowledges support from the Spanish State Research Agency project AYA2016-79425-C3-2-P.

\newpage

\appendix

\begin{figure}[ht!]
    \centering
    \includegraphics[width=0.98\textwidth]{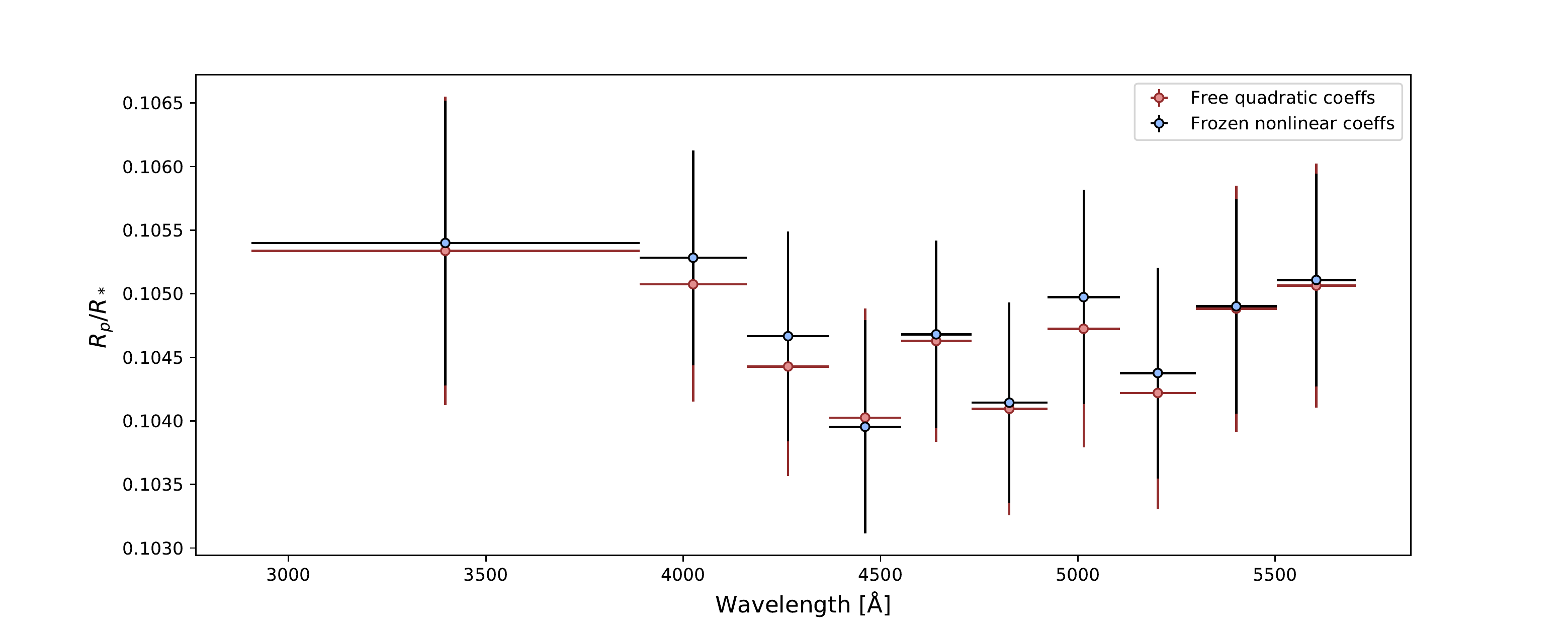}
    \caption{Comparison of transmission spectra from the G430L dataset obtained by two different treatments of limb darkening. Red points are inferred by fitting with a quadratic limb darkening law, with coefficients allowed to vary, and blue/black points are inferred by using the nonlinear limb darkening law, as is done in the main text (see Section~\ref{sec:limb_darkening}).}
    \label{fig:qd_vs_nl_g430l}
\end{figure}

\begin{figure}[ht!]
    \centering
    \includegraphics[width=0.98\textwidth]{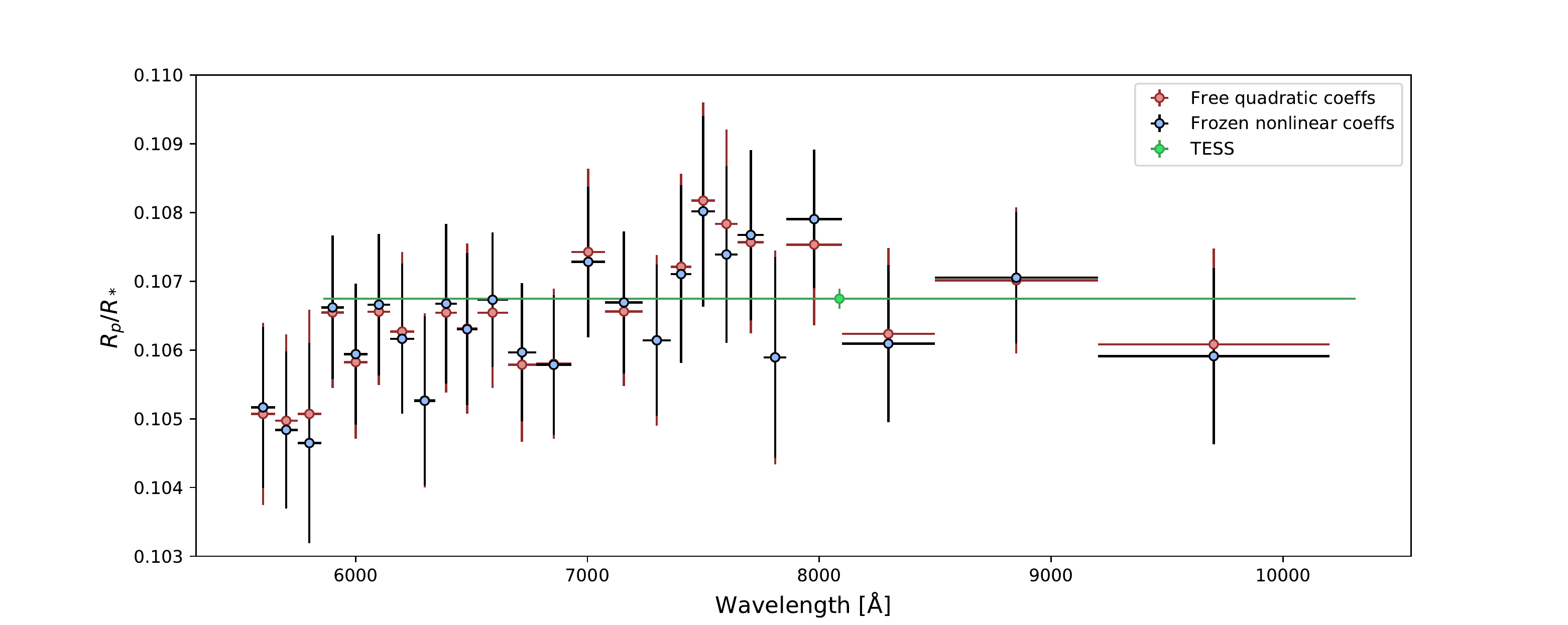}
    \caption{Same as Figure \ref{fig:qd_vs_nl_g430l}, but for the G750L grating.}
    \label{fig:qd_vs_nl_g750l}
\end{figure}

Here we demonstrate that different treatments of limb darkening only have a minor impact on the resulting transmission spectrum (Figure~\ref{fig:qd_vs_nl_g430l} and \ref{fig:qd_vs_nl_g750l}). We also include tabulated values for the transmission spectrum that is presented in the main text (Table~\ref{tab:tsres79}). 

\clearpage

\startlongtable
\setlength{\tabcolsep}{18pt}
\begin{deluxetable}{cccccc}
\tablecaption{Results of the Spectrophotometric Light Curve fits for WASP-79b\label{tab:tsres79}}
\tablehead{
\colhead{Wavelength Range [\text{\AA}]} & \colhead{$\text{R}_\text{p}/\text{R}_\star$} & \colhead{$\text{u}_\text{1}$} & \colhead{$\text{u}_\text{2}$} & \colhead{$\text{u}_\text{3}$} & \colhead{$\text{u}_\text{4}$} }
\startdata
2905 - 3890  &  0.1054 $\pm$ 0.0011  &  -0.1442  &  1.0796  &  -0.0603 &  -0.0976 \\
3890 - 4160  &  0.1053 $\pm$ 0.0008  &  -0.0435  &  0.4350  &  1.0150  &  -0.6323 \\
4160 - 4370  &  0.1047 $\pm$ 0.0008  &  -0.1230  &  0.7619  &  0.5932  &  -0.4646 \\
4370 - 4550  &  0.1040 $\pm$ 0.0008  &  -0.0330  &  0.4152  &  1.0049  &  -0.6321 \\
4550 - 4730  &  0.1047 $\pm$ 0.0007  &  -0.0793  &  0.6712  &  0.6262  &  -0.4770 \\
4730 - 4920  &  0.1041 $\pm$ 0.0008  &  -0.1129  &  0.9301  &  0.1773  &  -0.2873 \\
4920 - 5105  &  0.1050 $\pm$ 0.0008  &  -0.0565  &  0.6682  &  0.4954  &  -0.4065 \\
5105 - 5300  &  0.1044 $\pm$ 0.0008  &   0.0123  &  0.4235  &  0.7403  &  -0.4965 \\
5300 - 5505  &  0.1049 $\pm$ 0.0008  &  -0.0230  &  0.6028  &  0.4538  &  -0.3709 \\
5505 - 5705  &  0.1051 $\pm$ 0.0008  &  -0.0009  &  0.5412  &  0.4922  &  -0.3842 \\
5265 - 5550  &  0.1043 $\pm$ 0.0011  &  -0.0702  &  0.7920  &  0.2109  &  -0.2691 \\
5550 - 5650  &  0.1052 $\pm$ 0.0012  &  -0.0767  &  0.8429  &  0.1091  &  -0.2256 \\
5650 - 5750  &  0.1048 $\pm$ 0.0011  &  -0.0587  &  0.7875  &  0.1450  &  -0.2348 \\
5750 - 5850  &  0.1046 $\pm$ 0.0015  &  -0.0842  &  0.9007  &  -0.0084  &  -0.1737 \\
5850 - 5950  &  0.1066 $\pm$ 0.0010  &  -0.0857  &  0.9190  &  -0.0588  &  -0.1494 \\
5950 - 6050  &  0.1059 $\pm$ 0.0010  &  -0.0854  &  0.9270  &  -0.0982  &  -0.1270 \\
6050 - 6150  &  0.1067 $\pm$ 0.0010  &  -0.0275  &  0.7034  &  0.1682  &  -0.2350 \\
6150 - 6250  &  0.1062 $\pm$ 0.0011  &  -0.0831  &  0.9342  &  -0.1711  &  -0.0841 \\
6250 - 6345  &  0.1053 $\pm$ 0.0012  &  -0.0080  &  0.6429  &  0.2037  &  -0.2447 \\
6345 - 6435  &  0.1067 $\pm$ 0.0012  &  -0.0921  &  0.9854  &  -0.2515  &  -0.0544 \\
6435 - 6525  &  0.1063 $\pm$ 0.0011  &  -0.1001  &  1.0355  &  -0.3422  &  -0.0155 \\
6525 - 6655  &  0.1067 $\pm$ 0.0010  &  -0.1402  &  1.2937  &  -0.7606  &  0.1507 \\
6655 - 6780  &  0.1060 $\pm$ 0.0010  &  -0.0951  &  1.0179  &  -0.3571  &  -0.0022 \\
6780 - 6930  &  0.1058 $\pm$ 0.0010  &  -0.0955  &  1.0224  &  -0.3807  &  0.0092 \\
6930 - 7075  &  0.1073 $\pm$ 0.0011  &  -0.0975  &  1.0373  &  -0.4213  &  0.0281 \\
7075 - 7240  &  0.1067 $\pm$ 0.0010  &  -0.0972  &  1.0409  &  -0.4537  &  0.0457 \\
7240 - 7360  &  0.1061 $\pm$ 0.0011  &  -0.0979  &  1.0468  &  -0.4799  &  0.0587 \\
7360 - 7450  &  0.1071 $\pm$ 0.0013  &  -0.0978  &  1.0470  &  -0.4962  &  0.0674 \\
7450 - 7550  &  0.1080 $\pm$ 0.0014  &  -0.1008  &  1.0634  &  -0.5225  &  0.0771 \\
7550 - 7650  &  0.1074 $\pm$ 0.0013  &  -0.1000  &  1.0641  &  -0.5410  &  0.0875 \\
7650 - 7760  &  0.1077 $\pm$ 0.0012  &  -0.1015  &  1.0760  &  -0.5719  &  0.1014 \\
7760 - 7860  &  0.1059 $\pm$ 0.0015  &  -0.1026  &  1.0776  &  -0.5798  &  0.1048 \\
7860 - 8095  &  0.1079 $\pm$ 0.0010  &  -0.1047  &  1.0889  &  -0.6099  &  0.1182 \\
8095 - 8500  &  0.1061 $\pm$ 0.0011  &  -0.1161  &  1.1588  &  -0.7529  &  0.1802 \\
8500 - 9200  &  0.1071 $\pm$ 0.0010  &  -0.1197  &  1.1574  &  -0.7742  &  0.1880 \\
9200 - 10200 &  0.1059 $\pm$ 0.0013  &  -0.1009  &  1.0490  &  -0.6456  &  0.1358 \\
\enddata
\tablecomments{These are the values obtained after applying the stitching correction described in Section~\ref{subsec:binned_fits}.}
\end{deluxetable}

\bibliography{WASP79b.bib}{}
\bibliographystyle{aasjournal}

\end{document}